\documentstyle[12pt]{article}

\newcommand{\be}{\begin{equation}}
\newcommand{\ee}{\end{equation}}
\newcommand{\bea}{\begin{eqnarray}}
\newcommand{\eea}{\end{eqnarray}}

\begin{document}
\newcommand{\nd}[1]{/\hspace{-0.5em} #1}
\begin{titlepage}
\begin{flushright}
{\bf October 2003} \\ 
SWAT-384 \\ 
hep-th/0310117 \\
\end{flushright}
\begin{centering}
\vspace{.2in}
 {\large {\bf S-Duality, Deconstruction and Confinement for a} \\
{\bf Marginal Deformation of ${\cal N}=4$} {\bf SUSY Yang-Mills} 
}\\
\vspace{.4in}
Nick Dorey \\
\vspace{.4in}
Department of Physics, University of Wales Swansea \\
Singleton Park, Swansea, SA2 8PP, UK\\
\vspace{.2in}
%
%
\vspace{.4in}
{\bf Abstract} \\

\end{centering}
We study an exactly marginal deformation of ${\cal N}=4$ SUSY 
Yang-Mills with gauge group $U(N)$ using field theory and 
string theory methods. 
The classical theory has a Higgs branch for rational values of the 
deformation parameter. 
We argue that the quantum theory also has an S-dual  
confining branch which cannot be seen classically. 
The low-energy effective theory on 
these branches is a six-dimensional non-commutative 
gauge theory with sixteen supercharges. 
Confinement of magnetic and electric charges, on the Higgs and confining 
branches respectively, occurs due to 
the formation of BPS-saturated strings in the low energy theory.       
The results also suggest a new way of deconstructing 
Little String Theory as a large-$N$ limit of a 
confining gauge theory in four dimensions.  

\end{titlepage}
\section{Introduction}
\paragraph{}
Recent years have seen rapid progress in our understanding of 
supersymmetric (SUSY) gauge dynamics. In four dimensions, the 
best understood model is ${\cal N}=4$ SUSY Yang-Mills theory 
with gauge group $SU(N)$ which exhibits exact electric-magnetic duality and 
has a dual large-$N$ formulation as IIB string theory on 
$AdS_{5}\times S^{5}$. On the other hand, 
the ${\cal N}=4$ theory only has conformal and 
Coulomb phases, and thus has little in common with 
non-supersymmetric Yang-Mills theory in its confining phase. 
In this paper (and a subsequent one \cite{nd2})  
we will study a system which is as close as possible 
to the ${\cal N}=4$ theory and shares some of its special properties. 
Specifically, we will deform 
the ${\cal N}=4$ Lagrangian by adding an exactly marginal operator which 
preserves ${\cal N}=1$ supersymmetry. The resulting theory also exhibits many  
new features, including the emergence of extra dimensions and 
a confining phase with BPS-saturated strings. There are also 
interesting limits where the confining phase has a dual description in 
terms of six-dimensional Little String Theory \cite{LST}. 
We discuss similarities and differences between the 
marginally-deformed theory and  
the better understood case of a relevant deformation 
(the ${\cal N}=1^{*}$ theory). 
In particular, several phenomena which first emerged 
in Polchinski and Strassler's analysis of the ${\cal N}=1^{*}$ theory 
\cite{PS} will appear again here 
in a new and more tractable setting.                
\paragraph{}
In terms of ${\cal N}=1$ multiplets, the theory we will consider 
contains a single vector multiplet 
$V$ and three chiral multiplets ${\Phi}_{i}$ $i=1,2,3$ in the adjoint 
representation of the gauge group. The classical superpotential is, 
\begin{equation}
{\cal W}= i\kappa {\rm Tr}_{N}\left[
e^{i\frac{\beta}{2}}\Phi_{1}\Phi_{2}\Phi_{3}-e^{-i\frac{\beta}{2}}\Phi_{1}
\Phi_{3}\Phi_{2}\right]    
\label{LSsupx}
\end{equation}
The ${\cal N}=4$ theory is recovered by setting $\beta=0$ and $\kappa=1$. 
As we review in Section 2, there is known to be a 
surface of renormalization group fixed points 
in coupling-constant space which includes the ${\cal N}=4$ point \cite{LS1}.  
We will refer to the corresponding family of 
superconformal field theories collectively as the $\beta$-deformed theory. 
For earlier discussions of the finiteness of this model see \cite{early} 
and for other relevant work see \cite{D1}-\cite{DHK}.           
In this paper, we will mainly be interested in the $\beta$-deformed theory 
with gauge group $U(N)$. However we also discuss the $SU(N)$ theory, 
highlighting some important differences between the two cases.  
\paragraph{}
Both theories already exhibit some interesting features at 
the classical level. In particular, new Higgs branches appear 
when the deformation parameter $\beta/2\pi$ takes any real 
rational value (for appropriate values of $N$). 
We will focus on a particular Higgs branch of the $U(N)$ theory, 
denoted ${\cal H}_{m}$ which occurs when 
$\beta=2\pi/n$ with $N=mn$. 
At a generic point on this branch, the $U(N)$ gauge group 
is broken to $U(m)$ at a scale\footnote{In fact, the VEVs of the three complex scalars are independent on this branch and can provide more than one relevant scale. We discuss this more general case in Sections 3 and 4 below.} 
${\rm v}$ set by the scalar vacuum expectation 
value (VEV). A surprising aspect of the classical physics on this 
branch is the appearance of additional spacetime dimensions at 
low energy. In Section 4, we calculate the exact classical spectrum of the 
theory and show that it has a natural Kaluza-Klein interpretation in terms of 
a six-dimensional theory. Using the observations of \cite{AF1}, we can 
rewrite the action of the four-dimensional $U(N)$ theory as that of 
a six-dimensional $U(m)$ theory with two spatial dimensions 
discretized on an $n\times n$ lattice with periodic boundary conditions. 
More precisely, the theory is a lattice regularization 
of a six-dimensional non-commutative $U(m)$ gauge theory with 
${\cal N}=(1,1)$ supersymmetry. The theory has lattice spacing 
$\varepsilon \sim {\rm v}^{-1}$ and is compactified on a two-dimensional torus 
of radii $R\sim n\varepsilon$. On length-scales 
much larger than $\varepsilon$, the classical lattice action is 
well approximated by its continuum counterpart which is 
a six-dimensional ${\cal N}=(1,1)$ theory with gauge group $
U(m)$ compactified on the non-commutative torus, $T^{2}_{\Theta}$. The 
dimensionless non-commutativity parameter $\Theta$ takes the value $1/n$.  
\paragraph{}
This kind of equivalence between a 
lattice gauge theory and a lower-dimensional theory in its 
Higgs phase is known as deconstruction \cite{AHCG, AHCK} 
(see also \cite{Hal,Set}). The relation to the original 
version of deconstruction, which yields an ordinary commutative 
lattice gauge theory, was explained in \cite{AF1}. The appearance of 
extra supersymmetry at low-energies and at large-$N$ is a familar 
feature of deconstruction. In the present context it has a very prosaic 
explanation: the theory on ${\cal H}_{m}$ 
has deformation parameter $\beta=2\pi/n=2\pi m/N$ which goes to zero as 
$N\rightarrow \infty$ and we recover the ${\cal N}=4$ Lagrangian with 
sixteen supercharges for any fixed value of the gauge coupling. 
Note that we are expanding around a 
background which is not a vacuum of the ${\cal N}=4$ theory for any 
finite values $N$, but becomes so in the limit $N\rightarrow \infty$. 
The enhancement of supersymmetry allows us to argue that the classical 
spectrum of Kaluza-Klein states becomes exact in the 
$N\rightarrow \infty$ limit\footnote{Throughout this paper the 
$N\rightarrow \infty$ limit refers to the limit $n\rightarrow \infty$ with 
$m$ fixed and $N=mn$.}.       
\paragraph{}
The appearance of a non-commutative theory in six dimensions is also 
very reminiscent of the ${\cal N}=1^{*}$ theory. The latter theory also 
has a vacuum in which $U(N)$ is Higgsed down to $U(m)$, with $m|N$, 
although the resulting physics is quite different because 
the unbroken $U(m)$ is confined at low energies. The string theory 
realization of this ground state given in \cite{PS} involves 
$N$ D3 branes polarized \cite{Myers} by external 
Ramond-Ramond fields into $m$ D5 branes wrapped on a two-sphere (see also 
\cite{sr}). The D3 brane 
charge appears as a background $B_{\rm NS}$ two-form potential leading to a 
six-dimensional non-commutative $U(m)$ gauge theory on the D5 
world-volume. 
As discussed in Section 8 below, the Higgs branch ${\cal H}_{m}$ of 
the $\beta$-deformed theory has a very similar string theory realization in 
which the $m$ D5-branes are wrapped on a two-dimensional 
torus rather than a sphere. 
The different physics of the two cases 
can be traced to this distinction. 
Toroidal compactification with periodic boundary conditions preserves the 
full sixteen supercharges of the D5 brane theory 
(although the lattice regularization implied by finite $N$ does not). 
At energies far below the compactification scale, the $U(m)$ 
theory on the Higgs branch therefore 
reduces to a conformal ${\cal N}=4$ super Yang-Mills in four dimensions. 
In contrast, spherical compactification breaks most of the supersymmetry, 
leaving an ${\cal N}=1$ theory in four dimensions 
which becomes strongly-coupled and confining in the IR (for $m>1$).  
\paragraph{}
By choosing the UV coupling to be small  
we can ensure that the theory on the Higgs branch ${\cal H}_{m}$ 
is weakly-coupled for all energy scales. In this regime, the low-energy 
effective theory is the six-dimensional non-commutative gauge 
theory described above which is also weakly-coupled throughout its range 
of validity justifying a semiclassical analysis. 
In particular, one can find an explicit map between the fields 
of the original four-dimensional gauge theory and those of the 
six-dimensional effective theory. As we explain in Section 5, this map 
is a consequence of the standard Morita equivalence of gauge theory on the 
non-commutative torus $T^{2}_{\Theta}$.    
\paragraph{}
The low-energy effective theory has classical solutions corresponding to 
BPS-saturated strings. These solutions are obtained by embedding 
four-dimensional non-commutative $U(m)$ Yang-Mills instantons as static 
solutions in six dimensions. The resulting strings have tension 
$T\sim {\rm v}^{2}/g^{2}n$. 
In Section 5, we use the Morita equivalence of the 
low-energy theory on $T^{2}_{\Theta}$ to show that these 
solutions correspond to magnetic flux tubes in the original 
$U(N)$ gauge theory. As usual, the 
formation of finite-tension magnetic flux-tubes indicates the 
confinement of external magnetic charges. At large $N$, the 
core-size of the flux tube is always much larger than the lattice spacing 
justifying the use of the continuum low-energy theory.  
Using the Morita map, standard facts 
about non-commutative instantons translate into 
interesting predictions for the flux tubes of this theory. On the Higgs 
branch ${\cal H}_{m}$, where the unbroken gauge group is $U(m)$ the flux 
tubes have a variable core size corresponding to the scale-size of the 
instanton. Increasing the core-size allows the magnetic flux lines of the 
unbroken $U(m)$ to spread out. There are also interesting vacua where the 
gauge group is further broken down to $U(1)^{m}$. In these vacua we find 
BPS magnetic flux tubes of fixed core-size\footnote{For other recent 
work on non-abelian flux tubes in deformations of ${\cal N}=4$ SUSY Yang-Mills see \cite{Kn}}.          
\paragraph{}
Another remarkable feature of the $\beta$-deformed theory is that it has an 
exact $SL(2,{\bf Z})$ duality inherited from the S-duality of the underlying 
${\cal N}=4$ theory. The evidence for this duality is 
reviewed in Section 2 below. This includes a linearized analysis of the 
perturbation and also the exact effective superpotential for a 
massive version of the $\beta$-deformed theory derived in \cite{DHK}. 
The $SL(2,{\bf Z})$ generator corresponding to 
electric-magnetic duality relates the theory with coupling 
$\tau$ and deformation parameter $\beta=2\pi/n$ to a dual theory with coupling 
$\tilde{\tau}=-1/\tau$ and deformation parameter \footnote{This form of 
the transformation is only correct for $n>>1$. More generally $\tau$ 
recieves a finite algebraic renormalisation which is 
given in (\ref{taur}) below.} $\tilde{\beta}=2\pi\tilde{\tau}/n$. 
The electric Higgs branch 
${\cal H}_{m}$ in the former theory is mapped to a dual 
magnetic Higgs branch $\tilde{\cal H}_{m}$ in the latter. The magnetic branch 
occurs at a complex value of $\beta$ and does not coincide with any 
vacuum states of the classical theory. The other elements of 
$SL(2,{\bf Z})$ map the Higgs branch onto an infinite family of new branches 
realised in various oblique confining phases.  
The existence of these new branches 
can be demonstrated directly by taking an appropriate massless limit of 
the results in \cite{DHK}.  
\paragraph{}
The S-duality described above maps the weak-coupling limit of the  
electric Higgs branch theory to a dual regime of the magnetic Higgs branch   
theory. In the dual picture, the classical BPS strings of the 
six-dimensional effective theory correspond to electric flux-tubes 
which confine electric charges. The confinement is 
partial in the sense that it only applies objects which are charged under the 
broken generators of the magnetic gauge group.  
This corresponds to a phase where the original electric gauge group $U(N)$ 
is confined down to a $U(m)$ subgroup. We also discuss phases where 
$U(N)$ is further confined down to $U(1)^{m}$ which are S-dual 
to the Higgs phases with unbroken gauge group $U(1)^{m}$. 
The confining phase of the $\beta$-deformed theory 
is novel for several reasons.  
Apart from the appearance of extra dimensions and ${\cal N}=4$ supersymmetry 
at low energy, confinement occurs in a phase where conformal invariance is spontaneously broken leading to massless Goldstone bosons\footnote{Of course the coexistence of confinement and massless Goldstone modes is a well-known feature of QCD with massless quarks and is not, by itself, surprising}.    
Importantly, as the theory contains only adjoint fields, the Higgs and 
confining phases are genuinely distinct.      
\paragraph{}
In the semiclassical regime discussed above long strings appear as classical 
solitons in the low-energy theory. 
Thus the mass scale set by 
the string tension is large compared with the masses of elementary quanta. 
An extrapolation of the semiclassical formula for the 
tension indicates that there should also be other regimes 
where the confining strings become 
light. The fact that the strings become 
BPS saturated with respect to the enlarged supersymmetry as 
$N\rightarrow \infty$ suggests that this extrapolation should be 
reliable for large $N$. In the phase discussed above 
where the gauge group is confined to 
$U(1)^{m}$ and we have flux tubes of fixed core-size, one might expect an 
infinite tower of states 
corresponding to the excitations of the light string. 
This would match the expected behaviour  
of a large-$N$ confining gauge theory with adjoint fields, where the 
spectrum should include an infinite tower of glueball states. 
In Section 9 below, we will identify a regime of 
parameters where we believe that this behaviour can be exhibited 
explicitly.   
\paragraph{}
In Section 8, we reinforce some of our conclusions by studying a   
string-theory realisation of the $\beta$-deformed theory. 
As mentioned above, this involves $N$ D3 branes in the presence of 
background RR fields. The Higgs branch ${\cal H}_{m}$ 
corresponds to a configuration in which the D3-branes 
are polarized into $m$ D5 branes wrapped 
on a two-dimensional torus of radii $r\sim 2\pi \alpha' {\rm v}$.  
The presence of $N$ units of D3 brane charge leads to  
non-commutativity on the D5 brane world-volume. We  
check that the low-energy theory on the D5 world-volume precisely matches the 
six-dimensional effective action obtained in Section 4 by field 
theory methods. 
\paragraph{}
The string theory construction also provides a 
nice picture of the electric and magnetic confinement mechanisms 
discussed above. The magnetic flux tube is realised as a bound-state of a 
D-string with the D5 branes. This picture confirms the identification between 
six-dimensional instanton strings and 
four-dimensional magnetic flux tubes described above.  
The $SL(2,{\bf Z})$ duality of the $\beta$-deformed theory corresponds to the 
S-duality of IIB string theory. Performing an S-duality transformation 
on the toroidally wrapped D5 branes leads to a configuration of 
$m$ NS5-branes wrapped on the same torus. This is the string theory 
realisation of the theory on the magnetic Higgs branch denoted 
$\tilde{\cal H}_{m}$ above. The electric flux tubes of the confining 
phase are realised as fundamental strings bound to the NS5-branes.
This is very similar to the discussion of confinement 
in the ${\cal N}=1^{*}$ theory given by Polchinski and Strassler \cite{PS}. 
\paragraph{}
So far we have mainly discussed the semiclassical regime where 
the length-scale set 
by the six-dimensional gauge coupling is much smaller than the 
lattice spacing. As for any lattice model, 
an obvious question is whether the theory has an 
interacting continuum limit. It turns out that this is closely 
related to the issue raised above of the possible existence of 
regimes with light strings. In Section 9, we present a preliminary 
discussion of this question. We propose a one-parameter 
family of continuum limits. Via S-duality these limits can be interpreted as 
large-$N$ limits of the confining phase theory. We identify 
the resulting continuum theory as a decoupled theory on the NS5 branes 
appearing in the brane construction of Section 8. In one special case we 
obtain commutative Little String Theory in six non-compact dimensions.       
In addition, we identify an alternative limit 
analogous to the one considered in 
\cite{AHCK}, which yields commutative Little String Theory 
compactified to four 
dimensions on a torus of fixed size. Strikingly, the limit in 
question is simply a 
't Hooft limit of the $\beta$-deformed theory in a fixed vacuum state on 
its confining branch $\tilde{\cal H}_{m}$. Finally, we also 
identify field theory 
limits which are dual to double-scaled Little String Theory \cite{GK}. 
This case is particularly interesting as the full spectrum of the theory 
can be calculated explicitly in a particular regime of parameters. 
The results discussed in Section 9 are presented in more detail in 
\cite{nd2}. 
\paragraph{}
The rest of the paper is organised as follows. In Section 2, we introduce the 
$\beta$-deformed theory and discuss its conformal invariance and S-duality. 
In Section 3, we analyse the classical vacuum structure of the theory. 
Section 4 is devoted to the classical equivalence between the 
$\beta$-deformed theory and the six-dimensional lattice theory discussed 
above. In Section 5, we derive the classical low energy effective action and 
comment on quantum corrections. Section 6 discusses the classical physics of 
BPS strings on the Higgs branch. Section 7 presents evidence for the 
existence of the S-dual confining branches. In Section 8 we present a  
string theory construction of the $\beta$-deformed theory. 
The continuum limit and its relation to Little String Theory is considered in 
Section 9. 
         
\section{The ${\cal N}=4$ Theory and its Deformations}
\paragraph{}
We start by considering ${\cal N}=4$ SUSY Yang-Mills theory with 
gauge group $U(N)$ or $SU(N)$ and complexified coupling constant 
$\tau=4\pi i/g^{2} + \theta/2\pi$. The theory contains a single vector 
multiplet of ${\cal N}=4$ supersymmetry which includes the gauge field 
together with four species of adjoint Weyl fermions and six real 
adjoint scalars, which transform in the ${\bf 4}$ and ${\bf 6}$ of the 
$SU(4)$ R-symmetry group respectively. This theory has three remarkable but, 
by now, well-established properties: 
\paragraph{}
{\bf 1:} The $\beta$-function of the gauge coupling, $g^{2}$, is exactly 
zero. The theory is finite and, in the absence of scalar 
vacuum expectation values (VEVs), it is conformally invariant \cite{finite}. 
\paragraph{}
{\bf 2:} The theory has an exact S-duality \cite{MO} which acts on 
the complexified coupling as $\tau \rightarrow (a\tau + b)/(c \tau+ d)$ 
where the integers $a$, $b$, $c$ and $d$, with $ad-bc=1$ define an element 
of the modular group $SL(2,{\bf Z})$. 
\paragraph{}
{\bf 3:} In the 't Hooft large-$N$ limit, 
$N\rightarrow \infty$ with $\lambda=g^{2}N$ held fixed, 
the $SU(N)$ theory is dual to Type IIB superstring theory on $AdS_{5}\times 
S^{5}$ of radius ${\cal R}$, with ${\cal R}^{4}/\alpha'^{2}=\lambda$ 
\cite{Mal}. The dual theory can be reliably approximated by 
IIB supergravity for $\lambda>>1$.           
\paragraph{}
The basic objects in the conformally invariant ${\cal N}=4$ theory are 
operators, $\hat{O}$ of definite scaling dimension $\Delta$. 
As in any conformal theory, a natural operation is to deform the theory 
by adding one of these operators to the Lagrangian; 
\begin{equation}
{\cal L}={\cal L}_{{\cal N}=4}+\mu\hat{O}
\label{deform}
\end{equation}    
In four-dimensions, the operator is irrelevant if $\Delta>4$. 
In this case the perturbation grows in the UV and extra information is 
needed to define the theory at short-distances. On the other hand if 
$\Delta<4$, the operator is relevant and the perturbation grows in the IR. 
In this case the theory typically flows to another CFT with lower central 
charge or a massive theory at low energies. When $\Delta=4$, then 
$\hat{O}$ is marginal, and the deformation leads to a new 
conformal theory with the same central charge. In this case the corresponding 
coupling, $\mu$, is dimensionless with vanishing $\beta$-function and it 
can be chosen to be small at all energy scales.  
Only in this case, does it make sense to treat the deformation 
as a small perturbation around the ${\cal N}=4$ theory.
\paragraph{}
In the following we will focus on a deformation of the 
${\cal N}=4$ theory with gauge group $U(N)$ or $SU(N)$ 
which is known to be exactly marginal and also 
preserves ${\cal N}=1$ supersymmetry. 
In the language of ${\cal N}=1$ 
supersymmetry, the ${\cal N}=4$ theory contains a single vector multiplet 
$V$ and three chiral multiplets ${\Phi}_{i}$ $i=1,2,3$ in the adjoint 
representation of the gauge group.  
The classical superpotential for the 
chiral multiplets is, 
\begin{equation}
{\cal W}_{{\cal N}=4}= 
i{\rm Tr}_{N}\left(\Phi_{1}[\Phi_{2},\Phi_{3}]\right)               
\label{n4sup}
\end{equation}
The theory we will study in this paper is obtained by deforming the 
commutator which appears in the classical superpotential. 
Specifically we have,   
\begin{equation}
{\cal W}= i\kappa {\rm Tr}_{N}\left(\Phi_{1}[\Phi_{2},\Phi_{3}]_{\beta}\right) \label{LSsup}
\end{equation}
where, 
\begin{equation}
[\Phi_{i},\Phi_{j}]_{\beta}=\exp\left(i\frac{\beta}{2}\right)
\Phi_{i}\Phi_{j}- \exp\left(-i\frac{\beta}{2}\right)
\Phi_{j}\Phi_{i}
\label{deformed}
\end{equation}
We have also introduced a parameter $\kappa$ which controls the 
normalization of the scalar potential relative to the kinetic terms in the 
Lagrangian. 
The original ${\cal N}=4$ theory is recovered for $\kappa=1$ and $\beta=0$. 
\paragraph{}
For non-zero values of the deformation parameter $\beta$, the $SU(4)$ 
R-symmetry of the ${\cal N}=4$ theory is broken down to $U(1)^{3}$. Each 
$U(1)$ rotates the phase of one of the three complex chiral superfields. 
The deformation also leads to important differences between the 
$U(N)$ and $SU(N)$ theories. In the undeformed $U(N)$ theory the 
photon of the central $U(1)$ subgroup is part of an ${\cal N}=4$ 
multiplet which also includes degrees of freedom corresponding to the traces 
of the adjoint scalar and fermion fields. 
The whole multiplet is completely decoupled from 
the traceless $SU(N)$ fields. For $\beta\neq 0$, the central photon and 
its ${\cal N}=1$ superpartner still decouple, but the chiral multiplets 
${\rm Tr}_{N}\Phi_{i}$ for $i=1,2,3$ do not. The $U(N)$ theory is equivalent 
to the $SU(N)$ theory with couplings of order $|\beta|$ to these 
extra chiral multiplets together with a decoupled $U(1)$ vector multiplet.     \paragraph{}
The deformation of the superpotential involves two linearly independent 
operators,  
\begin{eqnarray}
{\hat O}_{1}={\rm Tr}_{N} \left(i\Phi_{1}[\Phi_{2},\Phi_{3}]\right)  & 
\qquad{} \qquad{} & {\hat O}_{2}={\rm Tr}_{N} 
\left(i\Phi_{1}\{\Phi_{2},\Phi_{3}\}\right) \nonumber \\ 
\label{ops}
\end{eqnarray} 
At linear order in the deformation parameters $\beta$ and $\kappa-1$, the 
resulting modification of the ${\cal N}=4$ Lagrangian involves the   
the supersymmetric descendents of these operators obtained by acting 
with two supercharges. Schematically we have,    
\begin{eqnarray}  
{\cal L} & = & {\cal L}_{{\cal N}=4} \, + \frac{1}{g^{2}}
(\kappa-1)\,Q^{2}\hat{O}_{1}+\frac{1}{g^{2}}\beta\, Q^{2}\hat{O}_{2}\,\, + \,\, {\rm h.c.} 
\nonumber \\
\label{def1}
\end{eqnarray}
Note that we have chosen to normalize the fields so that every term in the 
Lagrangian has an overall prefactor of $1/g^{2}$.  
\paragraph{}
Both operators appearing in (\ref{def1}) have classical mass dimension four. 
However, the corresponding couplings 
are only exactly marginal if the operators 
do not get anomalous dimensions. 
At linear order in the deformation parameters, 
the dimensions of the operators appearing in (\ref{def1}) 
can be replaced by their ${\cal N}=4$ values. 
In the ${\cal N}=4$ theory, the operator $\hat{O}_{2}$ is part of a chiral 
primary multiplet. The full multiplet consists 
of symmetric, traceless, third-rank tensors of $SO(6)$ of the form 
$Tr_{N}(\phi^{\{I}\phi^{J}\phi^{K\}})$, where $\phi^{I}$ with 
$I=1,2,\ldots 6$ are the six real adjoint fields of the ${\cal N}=4$ theory\footnote{The complex adjoint scalars $\Phi_{i}$, $i=1,2,3$, appearing in the 
${\cal N}=1$ formulation of the theory can be written as 
$\Phi_{i}=\phi_{i}+ i\phi_{i+3}$.} \cite{Wads}. 
This means that neither $\hat{O}_{2}$ nor its descendent $Q^{2}\hat{O}_{2}$ 
acquire an 
anomalous dimension for any value of the ${\cal N}=4$ coupling $\tau$. 
On the other hand, $\hat{O}_{1}$, which involves an anti-symmetrized trace, 
is part of a non-chiral multiplet and its scaling dimension is not protected. 
In fact, the AdS/CFT correspondence relates these operators to stringy 
excitations which decouple in the SUGRA limit 
$\lambda\rightarrow \infty$. We therefore expect $\hat{O}_{1}$ and its 
descendent $Q^{2}\hat{O}_{1}$ to 
acquire large positive anomalous dimensions of order $\lambda^{\frac{1}{4}}$ 
at strong coupling.            
\paragraph{}
The argument given above suffices to show that the coupling $\beta$ is 
exactly marginal at linear order in the deformation parameters 
while $\kappa-1$ is not\footnote{The corresponding anomalous 
dimension is positive definite because of the Bogomol'nyi bound, and 
hence the coupling $\kappa-1$ is marginally irrelevant}. To show that a 
marginal coupling survives beyond linear order is much harder because we can 
no longer rely on ${\cal N}=4$ supersymmetry. Remarkably,  
it is still possible to make progress using the much weaker constraints 
of ${\cal N}=1$ supersymmetry. Leigh and Strassler \cite{LS1} studied the exact 
NSVZ $\beta$-functions for the three couplings $\tau$, $\beta$ and $\kappa$. 
The existence of a conformal fixed-point requires that 
all three $\beta$-functions vanish. This leads to three complex 
equations for three unknowns which typically has at most isolated solutions. 
However, in the present case, the authors of \cite{LS1} 
found that two of the three equations are 
redundant allowing for a two-complex parameter families of solutions. 
Specifically they found that the the theory has a critical surface 
in coupling constant space defined by $\kappa=
\kappa_{cr}[\tau,\beta]$ which includes the ${\cal N}=4$ point $\beta=0$, 
$\kappa=1$. The explicit form of the function 
$\kappa_{cr}[\tau,\beta]$ is unknown beyond one-loop. However, for 
$\beta=0$ and $\kappa=1$ we must recover the ${\cal N}=4$ theory. The 
linearized analysis described above indicates that near this point
the two marginal couplings are simply $\tau$ and $\beta$. 
Thus we have $\kappa_{cr}[\tau,\beta]=1+ O(|\beta|^{2})$.   
\paragraph{}
The resulting surface of complex dimension 
two is parametrized by complex couplings which are exactly marginal to 
all orders in perturbation theory and also non-perturbatively. 
It therefore corresponds to a two-parameter family of 
${\cal N}=1$ superconformal theories. As above, we will refer to these 
collectively as the $\beta$-deformed theory. In the following, we will rely 
heavily on the fact that the the theory has a smooth $\beta\rightarrow 0$ 
limit in which ${\cal N}=4$ supersymmetry is recovered. Note that one 
would not expect the corresponding limit to exist for a relevant 
deformation of the ${\cal N}=4$ theory. The 
$\beta$-deformation is not quite the most 
general possible marginal deformation of the ${\cal N}=4$ theory. 
It is also possible to add 
a term of the form $\rho Tr_{N}(\Phi_{1}^{3}+\Phi_{2}^{3}+\Phi_{3}^{3})$ 
to the superpotential leading to a three complex dimensional critical surface 
$\kappa=\kappa_{cr}[\tau,\beta,\rho]$. We will set $\rho=0$ from now on.
\paragraph{}
When the 't Hooft coupling is small, which corresponds to $g^{2}<<1/N$, 
we can study the 
$\beta$ deformation of the ${\cal N}=4$ theory using semiclassical methods. 
In the following (and in \cite{nd2}), 
we will also be interested in the behaviour of the 
deformed theory at strong coupling. When $1/N<< g^{2} << 1$, 
the undeformed ${\cal N}=4$ theory with gauge group $SU(N)$ is well 
approximated by IIB supergravity on $AdS_{5}\times S^{5}$. An obvious 
question is how the $\beta$-deformation is realised on the IIB side of the 
correspondence. As above, it is easiest to work to linear order in the 
deformation parameters 
where the exactly marginal coupling is $\beta$ itself. 
The corresponding operator 
$Q^{2}O_{2}$ is a supersymmetric decendent of a chiral primary and is 
therefore mapped to a supergravity field in $AdS_{5}$. 
\paragraph{}
The field in question is a particular $S^{5}$ spherical harmonic of the   
the IIB complexified three-form field strength, 
\begin{equation}
G_{(3)}=F_{(3)}-\tau H_{(3)}
\end{equation}
which transforms in the ${\bf 45}$ of $SU(4)\simeq SO(6)$. 
Turning on the $\beta$-deformation corresponds to introducing a source for 
this mode on the boundary of $AdS_{5}$. The source results in a small 
deformation of the $AdS_{5}\times S^{5}$ background which can be constructed 
by solving the supergravity field equations order by order in $|\beta|$. 
This program was carried out to second order in the deformation in 
\cite{AKY}. This is in marked contrast to the case of a relevant deformation, 
where the corresponding boundary source leads to a deformation 
of the geometry which grows in the interior of $AdS_{5}$ and 
cannot be treated perturbatively. 
As usual, the dual supergravity description becomes exact in the 
large-$N$ limit with $g^{2}$ fixed. The existence of a smooth 
supergravity description of the $\beta$-deformation will be important 
for what follows because it 
indicates that the $\beta\rightarrow 0$ limit commutes with 
this large-$N$ limit. 
\paragraph{} 
Another regime in which we can hope to make progress is that of 
ultra-strong coupling $g^{2}>>N$. In the undeformed case, the 
ultra strongly-coupled ${\cal N}=4$ theory is related to the weakly-coupled 
theory by an electric-magnetic duality transformation. 
We can also use S-duality to map an ${\cal N}=4$ theory with $1<< g^{2}<<N$,  
to one with $1/N<<g^{2}<<1$ which is well described by IIB supergravity.  
Thus we need to understand how the $SL(2,{\bf Z})$ duality properties 
of the ${\cal N}=4$ theory extend to the $\beta$-deformed case. 
Once again it is easiest to make progress at linear order in the deformation, 
where we are perturbing the ${\cal N}=4$ theory by adding the term 
$(\beta/g^{2})Q^{2}\hat{O}_{2}$ (and its hermitian conjugate) to the 
Lagrangian. In general we do not expect the new term in the 
Lagrangian to be invariant under the S-duality of the ${\cal N}=4$ theory. 
However, we will see that S-duality can be restored by assigning an 
appropriate transformation property to the complex coupling, $\beta$.    
\paragraph{}
A key property of the chiral primary operators of the ${\cal N}=4$ theory 
(and their supersymmetric descendents)  
is that they have well-defined transformation properties under 
$SL(2,{\bf Z})$ \cite{Int} (see also \cite{ADK}).  
In particular they transform as non-holomorphic modular forms. A modular 
form $F$ of weight $(w,\bar{w})$ transforms as, 
\begin{eqnarray}
F & \rightarrow & (c\tau+d)^{w}(c\bar{\tau}+d)^{\bar{w}}\, F     
\label{transform}
\end{eqnarray}
under the $SL(2,{\bf Z})$ transformation defined by integers $a$, $b$, $c$, $d$ with 
$ad-bc=1$. The basic rule is that a chiral primary of mass dimension $d$ 
obtained as a symmetrized trace of a product of $d$ scalars has 
weight $(\frac{d}{2},\frac{d}{2})$. Thus $\hat{O}_{2}$ has weight 
$(\frac{3}{2},\frac{3}{2})$. 
Following the conventions of \cite{ADK}, 
the left-handed supercharges $Q_{\alpha}$ have weights 
$(\frac{1}{4},-\frac{1}{4})$ and the gauge coupling $g^{2}$ has weight 
$(1,1)$. Thus we see that the S-duality of the ${\cal N}=4$ theory can be 
preserved in the presence of the deformation provided we assign $\beta$ 
holomorphic weight $(-1,0)$. In other words, 
at linear order in the deformation, 
we have the $SL(2,{\bf Z})$ transformation property, 
\begin{eqnarray}  
\beta \rightarrow \frac{\beta}{(c\tau+d)}  \nonumber \\
\label{sl2zlinear}
\end{eqnarray}    
\paragraph{}
So far our considerations are only valid for an infinitessimal deformation 
of the ${\cal N}=4$ theory. Beyond linear order, the 
$\beta$-deformation involves other operators beside $Q^{2}\hat{O}_{2}$ and 
we cannot even assume that this operator retains the same transformation 
properties it has in the ${\cal N}=4$ theory. As in the discussion of the 
existence of exactly marginal operators given above, we must now rely 
only on the much weaker constraints of ${\cal N}=1$ supersymmetry. 
Despite this, we now have strong evidence that S-duality does extend to 
all values of the marginal couplings. As we now review, one can even 
specify the exact transformation properties of these couplings. 
\paragraph{}
As we have seen, the $\beta$-deformation of the ${\cal N}=4$ 
theory leads to a two-parameter family of 
superconformal field theories. We can also further deform the 
theory by adding relevant operators. 
The resulting model then flows to the $\beta$-deformed conformal 
theory in the UV, but its behaviour will be different in the IR.  
In \cite{DHK}, a particular relevant deformation was introduced by 
adding an additional term, 
\begin{equation}
\Delta{\cal W}= {\rm Tr}_{N}
\left[ M\Phi_{1}\Phi_{2}+ \mu \Phi_{3}^{2}\right]   
\label{rel}
\end{equation}
to the superpotential (\ref{LSsup}) of the $\beta$-deformed theory. 
For $\beta=0$, the resulting theory is equivalent (by a global 
$SO(3)$ rotation) to the standard ${\cal N}=1^{*}$ deformation of 
${\cal N}=4$ supersymmetric Yang-Mills theory. 
The vacuum structure of the theory is well understood in this case 
\cite{VW, DW, ND} and exhibits an exact 
$SL(2,{\bf Z})$ duality, acting on the complex coupling $\tau$, 
which is inherited from the ${\cal N}=4$ theory. 
\paragraph{}
The ${\cal N}=1^{*}$ theory 
has a variety of vacuum states in different massless and 
massive phases. The massive vacua are realised in various Higgs and 
confining phases associated with the condensation of different 
electric and magnetically charged states. As the states which condense 
transform non-trivially under electric-magnetic duality, $SL(2,{\bf Z})$ acts 
by permuting different vacuum states \cite{DW}. For example the 
`S' generator of 
$SL(2,{\bf Z})$ interchanges vacua in the Higgs and confining phases.  
The full action of $SL(2,{\bf Z})$ on the ${\cal N}=1^{*}$ 
vacua is determined by the 
effective superpotential of the theory derived in \cite{ND}.  
The superpotential also determines the condensates of ${\cal N}=1$ 
chiral operators in each vacuum. For example, in any supersymmetric vacuum 
state we have,
\begin{equation}
\langle {\rm Tr}_{N} \Phi_{3}^{2} \rangle = 
\frac{\partial}{\partial \mu} \, \langle {\cal W}_{\rm eff}\rangle 
\label{u2}
\end{equation}
Similar results for other 
condensates can also be obtained \cite{DS}. The resulting 
formulae are consistent with the known modular weights of the same 
operators in the ${\cal N}=4$ theory. For example, the vacuum expectation 
value $\langle {\rm Tr }_{N}\Phi_{3}^{2} \rangle$ transforms 
like a modular form of weight $(1,1)$ modulo 
the vacuum permutations described above \cite{ADK}. 
\paragraph{}
Our ability to calculate exactly in the ${\cal N}=1^{*}$ theory is limited 
to holomorphic quantities like the chiral condensates described above. 
Thus we are only able to check the proposed $SL(2,{\bf Z})$ duality for these 
special quantities. However, the ${\cal N}=1^{*}$ theory flows in the UV 
to ${\cal N}=4$ theory where the corresponding 
duality is believed to be exact. As both the UV behaviour of the 
${\cal N}=1^{*}$ theory and its IR vacuum structure are $SL(2,{\bf Z})$ 
invariant, it is natural to conjecture that the duality is an exact 
property of the theory valid at all length-scales. Futher support for this 
view-point comes from the AdS/CFT correspondence where the $SL(2,{\bf Z})$ 
duality of the ${\cal N}=4$ theory is mapped to the 
S-duality of IIB string theory which is also believed to be exact. AdS 
duals of the massive ${\cal N}=1^{*}$ vacua were constructed in 
\cite{PS}. In particular, different vacua of the ${\cal N}=1^{*}$ theory were 
realised as different string backgrounds permuted by the exact S-duality 
of the IIB theory.                       
\paragraph{}
In \cite{DHK}, we presented evidence that a similar story is realised in 
the $\beta$-deformed case. Specifically, the exact effective 
superpotential of both the 
$SU(N)$ and $U(N)$ theories was determined via the Dijkgraaf-Vafa proceedure 
\cite{DV} by solving a related matrix model. In both case the results revealed 
an $SL(2,{\bf Z})$ duality acting on a renormalized gauge coupling, 
\begin{equation}
\tau_{R}=\frac{4\pi i}{g^{2}_{R}}+\frac{\theta_{R}}{2\pi}
=\tau+ \frac{iN}{\pi}\log\kappa
\label{taur}
\end{equation}
and also on the deformation parameters as
\begin{equation}
\tau_{R}\rightarrow \frac{a\tau_{R}+b}{c\tau_{R}+d}  \qquad{} \qquad{} 
\beta \rightarrow 
\frac{\beta}{c\tau_{R}+d}  \qquad{} \qquad{} \kappa^{2}\sin\beta \rightarrow 
\frac{\kappa^{2}\sin\beta}{c\tau_{R}+d}
\label{sl2z}
\end{equation}
although there are some complications for the 
$U(N)$ theory which we discuss in Appendix A. 
It is easy to check this reduces to standard S-duality at the ${\cal N}=4$ 
point $\beta=0$, $\kappa=1$ and also reproduces the result 
(\ref{sl2zlinear}) of the linearized analysis, namely that $\beta$ has 
modular weight $(-1,0)$. 
The results of \cite{DHK} show that chiral operators transform with the 
same modular weights as they do in the ${\cal N}=4$ theory. 
Like the S-duality of the ${\cal N}=4$ theory, 
we will assume that (\ref{sl2z}) is an exact duality of the $\beta$-deformed 
theory for all values of the parameters.    
      
\section{Classical Vacuum Structure}
\paragraph{}       
In this section we will study the classical vacuum structure of the 
$U(N)$ theory and also make some comments about quantum corrections. 
The F- and D-flatness conditions read, 
\begin{equation} 
[\Phi_{1},\Phi_{2}]_{\beta}=
[\Phi_{2},\Phi_{3}]_{\beta}=[\Phi_{3},\Phi_{1}]_{\beta}=0
\label{fflat}
\end{equation}
and 
\begin{equation} 
\sum_{i=1}^{3}\,[\Phi_{i},\Phi_{i}^{\dagger}]=0
\label{dflat}
\end{equation} 
respectively. For the ${\cal N}=4$ case, $\beta=0$, 
the deformed commutators appearing 
in the F-term constraint revert to ordinary ones. In this case the vacuum 
equations are solved by diagonalizing each of the three complex scalars, 
\begin{equation}
\langle \Phi_{i} \rangle =  {\rm Diag}\left[ \lambda^{(i)}_{1},
\lambda^{(i)}_{2},\ldots,\lambda^{(i)}_{N}\right]
\label{arb}
\end{equation}
The $3N^{2}$ complex eigenvalues 
$\lambda^{(i)}_{a}$, for $i=1,2,3$ and $a=1,2,\ldots,N$, are unconstrained. 
After taking into account the Weyl group, we recover 
the familiar Coulomb branch of the ${\cal N}=4$ theory. On this branch the 
$U(N)$ gauge symmetry is spontaneously broken down to its Cartan subalgebra 
$U(1)^{N}$ and he vacuum manifold is the symmetric product 
${\rm Sym}_{N}\, {\bf C}^{3}$.  
\paragraph{}
Introducing a generic, non-zero value of $\beta$ changes things considerably. 
The F-flatness conditions are no longer solved by arbitrary diagonal matrices 
(\ref{arb}). For each value of the Cartan index 
$a\in \{1,2,\ldots,N\}$, at most one of the three eigenvalues,  
$\lambda^{(1)}_{a}$, $\lambda^{(2)}_{a}$ and $\lambda^{(3)}_{a}$ can be 
non-zero. In the simplest case of gauge group $U(1)$, 
the Coulomb branch of the 
${\cal N}=4$ theory, which has three complex dimensions, is partially 
lifted leaving three 
complex lines which intersect at the origin. For $G=U(N)$ with $N>1$,  
the Coulomb branch is formed by taking an $N$-fold symmetric product in the 
usual way. 
\paragraph{}
If $\beta/2\pi$ is not a rational real number, 
the Coulomb branch vacua described above are 
the only classical supersymmetric groundstates for any value of $N$. 
To illustrate 
the new possibilities which occur at rational values of $\beta/2\pi$ 
we will focus on a particular one-parameter family of vacua on the 
Coulomb branch. We choose 
$\langle \Phi_{1} \rangle=\alpha_{1}U_{(N)}$ and $\langle \Phi_{2} \rangle=
\langle \Phi_{3} \rangle=0$, where $\alpha_{1}$ is a complex number and 
we define the $N\times N$ `clock' matrix, 
\begin{equation}
U_{N}={\rm Diag}\left[\omega_{N},\omega^{2}_{N},\ldots,\omega^{N-1}_{N},
\omega_{N}^{N}\right]
\label{clock}
\end{equation}
where $\omega_{N}=\exp(2\pi i/N)$ is an $N\,$'th root of unity. 
We denote this one-dimensional complex submanifold of the Coulomb branch 
${\cal C}_{1}$. 
Starting at the ${\cal N}=4$ point, $\beta=0$, 
the adjoint Higgs mechanism yields the familiar spectrum of 
$W^{\pm}$ gauge bosons and their superpartners. These states lie in 
$N^{2}$ BPS-saturated vector multiplets of  
${\cal N}=4$ supersymmetry with masses, 
\begin{equation}
M^{(ab)}= 2|\alpha_{1}|\sin\left(\frac{\pi}{N}|a-b|\right)
\label{mass1} 
\end{equation}
for $a,b=1,2,\ldots,N$. As these states are BPS saturated, 
this mass formula is exact.  
Turning on a non-zero value of $\beta$ breaks ${\cal N}=4$ supersymmetry and 
each ${\cal N}=4$ multiplet splits into ${\cal N}=1$ multiplets. 
In particular each massive BPS-saturated vector multiplet of 
${\cal N}=4$ SUSY splits into a massive vector multiplet of 
${\cal N}=1$ SUSY and two additional massive chiral multiplets. 
While the vector multiplet masses are unaffected, 
the chiral multiplet masses depend explicitly on $\beta$ as,   
\begin{equation}
M_{\pm}^{(ab)}= 2|\alpha_{1}|\sin\left(\left|\frac{\pi}{N}(a-b) 
\pm \frac{\beta}{2}\right|\right)
\label{mass2} 
\end{equation}
The unbroken ${\cal N}=1$ supersymmetry algebra does not have a 
central charge and so none of these states are BPS. This means that, 
for $\beta\neq 0$, the classical mass formulae (\ref{mass1},\ref{mass2}) 
can recieve quantum corrections. As the masses are protected in the 
${\cal N}=4$ limit, we expect that these corrections are small for 
$|\beta|<<1$. 
\paragraph{}
The classical mass formula (\ref{mass2}) indicates that new massless states 
occur when $\beta=2\pi l/N$ where $l$ is an integer. The occurence of 
these states reflects the presence of a new branch of vacua. We begin by 
focussing on the case $l=1$. When $\beta=2\pi/N$, the F-term equations have 
non-trivial solutions involving the `clock' matrix $U_{(N)}$ introduced above 
and the $N\times N$ 'shift' matrix $V_{(N)}$ defined by, 
\begin{eqnarray}
\left(V_{(N)}\right)_{ab} & \qquad{}  = \qquad{}  & 1 \qquad{} 
{\rm if}\,\, b=a+1 \,\,\, {\rm mod}\,N \nonumber \\
& \qquad{} = \qquad{} & 0 \qquad {\rm otherwise} \nonumber \\
\label{shift}
\end{eqnarray}
These $N\times N$ matrices are both unitary and satisfy, 
\begin{equation}
\left[ U_{(N)},V_{(N)}\right]_{\frac{2\pi}{N}}=\sqrt{\omega_{N}} 
U_{(N)}V_{(N)} -
\sqrt{\bar{\omega}_{N}}V_{(N)}U_{(N)}=0
\label{hberg}
\end{equation}           
Using this relation we find a three complex parameter branch of vacua. Up to 
gauge transformations the scalar VEVs can be chosen as, 
\begin{equation}
\begin{array}{ccc} \langle \Phi_{1}\rangle = \alpha_{1}U_{(N)} & \qquad{} 
\qquad{} 
\langle  \Phi_{2}\rangle = \alpha_{2}V_{(N)} & \qquad{} \qquad{}  
\langle \Phi_{3}\rangle = \alpha_{3}W_{(N)} \end{array}
\end{equation}
where $W_{N}=V_{(N)}^{\dagger}U_{(N)}^{\dagger}$. Here $\alpha_{1}$, 
$\alpha_{2}$ and $\alpha_{3}$ are complex numbers. At a generic point on 
this branch of 
vacua, the $U(N)$ gauge symmetry is broken to its central $U(1)$ subgroup 
and the $U(1)^{3}$ R-symmetry is broken to 
${\bf Z}_{N}\times {\bf Z}_{N}$. As the expectation values of the 
adjoint scalars are traceless, this branch is also present in the 
$SU(N)$ theory. In the $SU(N)$ case the gauge group is completely broken. 
For future convenience, we will refer to this branch as the Higgs branch 
${\cal H}_{1}$. 
\paragraph{}
There is still some redundancy our parametrization of the Higgs branch. 
Specifically, the $U(N)$ gauge transformations; 
\begin{eqnarray}
\Phi_{i} & \rightarrow & \Gamma_{j}\Phi_{i}\Gamma^{\dagger}_{j} 
\nonumber \\
\label{ident}
\end{eqnarray}    
for $i=1,2,3$ and $j=1,2$ with $\Gamma_{1}=U_{(N)}$ and $\Gamma_{2}=V_{(N)}$ 
lead to discrete phase rotations,  
\begin{equation}
\begin{array}{cccc} \Gamma_{1}: \qquad{} & 
\alpha_{1} \rightarrow \alpha_{1} 
& \alpha_{2}\rightarrow \bar{\omega}_{N}\alpha_{2} & \alpha_{3} \rightarrow 
\omega_{N}\alpha_{3} \\  \Gamma_{2}: \qquad{} & 
\alpha_{1} \rightarrow \omega_{N}\alpha_{1} 
& \alpha_{2}\rightarrow \alpha_{2} & \alpha_{3} \rightarrow 
\bar{\omega}_{N}\alpha_{3}  \end{array}
\end{equation}
Taking into account these identifications the Higgs branch ${\cal H}_{1}$ 
is the complex orbifold ${\bf C}^{3}/{\bf Z}_{N}\times {\bf Z}_{N}$.   
This Higgs branch intersects the Coulomb branch on the submanifold 
${\cal C}_{1}$ defined above when $\alpha_{2}=\alpha_{3}=0$. The 
intersection is a fixed-line of the orbifold on which $\Gamma_{1}$ acts 
trivially. In addition, ${\cal H}_{1}$ intersects other components of 
the Coulomb branch when $\alpha_{1}=\alpha_{3}=0$, $\alpha_{2}\neq 0$ and 
when  $\alpha_{1}=\alpha_{2}=0$, $\alpha_{3}\neq 0$.
\paragraph{}
It will be useful to have a description of the vacuum structure in terms of 
gauge invariant variables. Hence we define the gauge invariant chiral 
operators,              
\begin{equation} 
u_{(k_{1},k_{2},k_{3})} = \frac{1}{N} {\rm Tr}_{N}
\left[\Phi_{1}^{k_{1}} \Phi_{2}^{k_{2}}\Phi_{3}^{k_{3}}\right]
\label{chiral}
\end{equation}
On the Higgs branch ${\cal H}_{1}$ discussed above, these VEVs of these 
operators vanish unless {\em either}
\begin{eqnarray}
(k_{1},k_{2},k_{3})=(0,0,0) \qquad{} {\rm mod}\,\,N & \qquad{} {\rm or} 
\qquad{} & k_{1}=k_{2}=k_{3} \qquad{} {\rm mod}\,\, N \nonumber \\
\label{chiralvevs1}
\end{eqnarray}
in which case, 
\begin{eqnarray}
\langle u_{(k_{1},k_{2},k_{3})}\rangle & \qquad{}  = \qquad{}  & 
\exp (i\nu_{(k_{1},k_{2},k_{3})})\alpha_{1}^{k_{1}}\alpha_{2}^{k_{2}}\alpha_{3}^{k_{3}} \nonumber \\
\label{chiralvevs2}
\end{eqnarray}
where $\exp(i\nu_{(k_{1},k_{2},k_{3})})$ is an unimportant 
phase. 
\paragraph{}
On this branch, the chiral ring is generated by four non-zero elements, 
$x=u_{(N,0,0)}$, $y=u_{(0,N,0)}$ 
$z=u_{(0,0,N)}$ and $w=u_{(1,1,1)}$ subject to the relation 
\begin{equation}
w^{N}=xyz
\label{orb1}
\end{equation} 
Equivalently we can think of $x$, $y$, $z$ and $w$ as gauge-invariant complex 
coordinates on the Higgs branch ${\cal H}_{1}$. 
The equation (\ref{orb1}) defines the complex orbifold 
${\bf C}^{3}/{\bf Z}_{N}\times {\bf Z}_{N}$. 
\paragraph{}
The massless modes on the Higgs branch ${\cal H}_{1}$ are the scalar moduli 
$\alpha_{i}$, $i=1,2,3$ and the photon of the central $U(1)$. 
Each of these bosonic fields is paired with a massless Weyl fermion by 
the unbroken ${\cal N}=1$ supersymmetry. These fields are free at low 
energies and the effective action is precisely that of an ${\cal N}=4$ 
supersymmetric gauge theory with gauge group $H=U(1)$. 
Note that this enhancement of 
supersymmetry in the IR theory is special to the $U(N)$ theory and 
does not occur for gauge group $SU(N)$. The complex 
coupling of the low-energy theory is related to that of the original theory 
as $\tau_{H}=N\tau$. Equivalently the low-energy 
$U(1)$ gauge coupling is $g_{H}^{2}=g^{2}/N$. 
As $\tau_{H}$ is an 
F-term coupling any corrections to the classical relation $\tau_{H}=N\tau$ 
must be holomorphic in $\tau$. The results of \cite{DHK}, show that the 
exact low energy coupling is $N\tau_{R}$, where the renormalised coupling 
$\tau_{R}$ is given in Eq (\ref{taur}) above.     
\paragraph{}
The $\beta$-deformed Lagrangian has ${\cal N}=1$ superconformal invariance 
and $U(1)^{3}$ R-symmetry. 
These symmetries are spontaneously broken on the 
Higgs branch and some of the massless scalars discussed above correspond 
to massless Goldstone bosons. The remaining massless modes lie in 
${\cal N}=1$ multiplets  with the Goldstone modes. 
As the symmetries in question are non-anomalous, 
all these fields remain massless in the full quantum theory. Thus we deduce 
that the Higgs branch ${\cal H}_{1}$ cannot be lifted by quantum effects.
The equation (\ref{orb1}), which defines the Higgs branch as a complex 
manifold is also protected from quantum correstions. For example, additional 
terms in (\ref{orb1}) 
which resolve the orbifold singularity are ruled out by the 
non-anomalous R-symmetries of the theory.    
\paragraph{}    
In the following, we will also be interested in more general Higgs branches 
of the theory with 
a larger unbroken gauge group. These occur when the rank $N$ has a non-trivial 
divisor. Thus we have $N=mn$ for some integers $m$ and $n$. If the deformation 
parameter takes the value $\beta=2\pi/n$ we find a three complex parameter 
branch on which the scalar VEVs can be chosen as,    
\begin{equation}
\begin{array}{ccc} \langle \Phi_{1}\rangle = \alpha_{1}I_{(m)}\otimes U_{(n)} & \qquad{}  
\langle  \Phi_{2}\rangle = \alpha_{2}I_{(m)}\otimes V_{(n)} & \qquad{}   
\langle \Phi_{3}\rangle = \alpha_{3}I_{(m)} \otimes W_{(n)} \end{array}
\label{vacm}
\end{equation}
where $I_{(m)}$ is the $m\times m$ unit matrix. Identifications generalising 
(\ref{ident}), lead to a Higgs branch ${\cal H}_{m}$ which is the complex 
orbifold ${\bf C}^{3}/{\bf Z}_{n}\times {\bf Z}_{n}$. On this branch, 
the VEVs of the operators $u_{(k_{1},k_{2},k_{3})}$ introduced above 
are given by replacing $N$ by $n$ in (\ref{chiralvevs1}) and 
(\ref{chiralvevs2}).  As for ${\cal H}_{1}$, 
the flat directions are protected by Goldstone's theorem and persist for 
all values of the gauge coupling. The low-energy theory on this branch is 
${\cal N}=4$ SUSY Yang-Mills theory with gauge group $H=U(m)$ 
with complexified coupling $\tau_{H}=\tau/n$. Equivalently, 
the low-energy gauge coupling is $g^{2}_{H}=g^{2}/n$. 
\paragraph{}
The low-energy theory on ${\cal H}_{m}$ is an ${\cal N}=4$ theory at its 
conformal point. This theory also has a Coulomb branch and it is quite 
straightforward to find the corresponding vacua of the original $U(N)$ 
theory. They are obtained by choosing the scalar VEVs as,    
\begin{equation}
\begin{array}{ccc} \langle \Phi_{1}\rangle = \alpha_{1}\Lambda^{(1)}\otimes 
U_{(n)} & \qquad{}  
\langle  \Phi_{2}\rangle = \alpha_{2}\Lambda^{(2)}\otimes V_{(n)} & \qquad{}   
\langle \Phi_{3}\rangle = \alpha_{3}\Lambda^{(3)} \otimes W_{(n)} \end{array}
\label{vac2}
\end{equation}
where $\Lambda^{(i)}$ $i=1,2,3$ are three diagonal $m\times m$ matrices, 
\begin{equation}
\Lambda^{(i)} =  {\rm Diag}\left[ \lambda^{(i)}_{1},
\lambda^{(i)}_{2},\ldots,\lambda^{(i)}_{m}\right]
\label{arb2}
\end{equation}
When these eigenvalues are distinct the unbroken gauge group is $U(1)^{m}$. 
This corresponds to the Coulomb phase of the low energy $U(m)$ theory 
described above. Apart from the Weyl group of $U(m)$ which permutes the 
eigenvalues we must also identify vacua which are related by discrete 
gauge transformations of the form (\ref{ident}). In the present case there 
are $2m$ such transformations generated by matrices 
$\Gamma_{1r}= P_{(r)}\otimes U_{(n)}$ and 
$\Gamma_{2r}= P_{(r)}\otimes V_{(n)}$ for $r=1,2,\ldots,m$ where 
$P^{(r)}_{st}=\delta_{sr}\delta_{tr}$. After dividing out by all of these 
transformations we find a vacuum manifold which is the symmetric product of 
$m$ copies of the orbifold ${\bf C}^{3}/{\bf Z}_{n}\times {\bf Z}_{n}$.
\paragraph{}
The appearance of the complex orbifold 
${\bf C}^{3}/{\bf Z}_{n}\times {\bf Z}_{n}$ naturally leads us to mention 
the standard string theory realisation of the $\beta$-deformed theory.  
As demonstrated in \cite{D1,D2}, the $U(N)$ theory 
with $N=mn$ and $\beta=2\pi/n$ arises on the world volume of 
$m$ D3 branes placed at a ${\bf C}^{3}/{\bf Z}_{n}\times{\bf Z}_{n}$ 
singularity with a single unit of {\it discrete torsion}. 
The discrete torsion corresponds to the weighting of the twisted sectors 
of the orbifold theory with the phase $\exp(i\beta)=\exp(2\pi i/n)$. 
The Higgs branches ${\cal H}_{m}$ considered above corresponds 
to the position of $m$ coincident D3-branes in the orbifold space. As usual, 
moving onto the Coulomb branch of the low energy theory corresponds to 
separating the branes. The resulting moduli space of vacua described 
in the previous paragraph is the symmetric product of $m$ copies 
of ${\bf C}^{3}/{\bf Z}_{n}\times {\bf Z}_{n}$. This corresponds to the 
full configuration space of $m$ D3 branes in the orbifold. 
This orbifold construction of the $\beta$-deformed theory is important 
for the discussion of deconstruction given in \cite{AF1}. 
We will also discuss a quite different stringy realisation of the Higgs branch 
in Section 8.       
\section{Classical Deconstruction}
\paragraph{}
In this Section we will discuss the $\beta$-deformed theory on the 
Higgs branch ${\cal H}_{m}$ 
and its Kaluza-Klein interpretation in terms of a 
six-dimensional lattice gauge theory. We will consider in turn the spectrum 
and the interactions. For simplicity we will restrict our attention 
to the submanifold ${\cal H}_{m}$ where 
only two of the three complex scalar fields acquire VEVs. Thus we set 
$\alpha_{3}=0$ in (\ref{vacm}). Correspondingly, we will adopt a convention 
where indices $i$, $j$ and $k$ take values $1$ or $2$ only. 
Throughout this section we will restrict our 
attention to the case of gauge group $U(N)$.  
\subsection{The Spectrum} 
\paragraph{}
To find the full classical spectrum in 
these vacua, we must expand the 
fields in fluctuations around their vacuum values and diagonalize the 
resulting quadratic terms in the action. To accomplish this we need to define 
a convenient basis which we now describe. 
\paragraph{}
All the fields of the theory are $N\times N$ matrices 
and we will think of them as 
elements of the algebra $gl(N, {\bf C})$. It will also be useful to use the 
decompose $gl(N,{\bf C})$ as a tensor product, 
\begin{equation}
gl(N,{\bf C}) \equiv gl(m,{\bf C})\otimes gl(n,{\bf C})        
\label{decomp1}
\end{equation}
We now choose the following set of $n^{2}$ 
basis elements for $gl(n,{\bf C})$ labelled by a 
vector $\vec{l}=(l_{1},l_{2})$ with integer components $l_{1}$ and $l_{2}$: 
\begin{equation}
J^{(\vec{l})}\, = \, V^{l_{1}}_{(n)}U_{(n)}^{-l_{2}}\, 
\omega_{n}^{\frac{l_{1}l_{2}}{2}}
\label{jl}
\end{equation}
Where $U_{(n)}$ and $V_{(n)}$ are $n\times n$ clock and shift matrices 
defined as in (\ref{clock}) and (\ref{shift}) above and 
$\omega_{n}=\exp(2\pi i/n)$. By virtue of the relations $U^{n}_{(n)}=
V_{(n)}^{n}=I_{(n)}$, the integers $l_{1}$ and $l_{2}$ are only defined 
modulo $n$. Equivalently, the vector $\vec{l}$ takes values in 
${\bf Z}_{n}^{2}$. The completeness of this basis means that for any 
$N\times N$ matrix $A$ we define the expansion,   
\begin{equation}
A=\frac{1}{n^{2}}\, 
\sum_{\vec{l} \in {\bf Z}_{n}^{2}}\, a^{(\vec{l})}\otimes 
J^{(\vec{l})}
\label{expan1}
\end{equation}
where each of the $n^{2}$ coefficients $a^{(\vec{l})}$ is 
itself an $m\times m$ matrix field. For a Hermitian matrix field 
$A(y)$, the additional constraint $a^{(\vec{l})\, \dagger}=a^{(-\vec{l})}$ must be imposed.  
\paragraph{}
We will begin by discussing the spectrum of massive gauge bosons on the 
Higgs branch ${\cal H}_{m}$. To this end we expand the $U(N)$ gauge field 
$A_{\mu}$ as in (\ref{expan1}). It is then easy to check that the 
corresponding coefficient $a_{\mu}^{(\vec{l})}$ is a mass eigenstate with 
eigenvalue,  
\begin{equation}
\left(M^{(\vec{l})}\right)^{2}= 
4|\alpha_{1}|^{2}\sin^{2}\left(\frac{l_{1}\pi}{n}\right) + 
4|\alpha_{2}|^{2}\sin^{2}\left(\frac{l_{2}\pi}{n}\right)
\label{mass3} 
\end{equation}
This result is consistent with the Higgs mechanism on ${\cal H}_{m}$ 
which breaks the gauge group $G=U(N)$ down to $H=U(m)$. In particular, 
the mode $a^{(\vec{0})}_{\mu}$ corresponds to the $m^{2}$ massless gluons 
of the low energy $U(m)$ gauge theory. The remaining $n^{2}-1$ modes
correspond to massive W-bosons and the degeneracy of $m^{2}$ states at each 
mass level reflects the fact that the W-bosons transform in the adjoint 
of the unbroken $U(m)$. 
\paragraph{}
As the theory has unbroken ${\cal N}=1$ supersymmetry, the full spectrum 
must consist only of ${\cal N}=1$ multiplets. The on-shell states 
which appear in each massive multiplet are classified according to their 
transformation properties under the $SU(2)$ little group. 
Restricting our attention to 
on-shell massive states with spin $\leq 1$, there are only two types of 
multiplet which can occur \cite{WB}. A massive 
vector multiplets contains one spin $1$ state, two spin $1/2$ states and 
one spin $0$ state. A massive chiral multiplet contains one spin $1/2$ 
state and two spin $0$ states. The spectrum of massive W-bosons found above 
must lie in $n^{2}-1$ massive vector multiplets each transforming in the 
adjoint of $U(m)$. The remaining states in the spectrum, 
having spins $\leq 1/2$, must be packaged into massive chiral 
multiplets. 
\paragraph{}
Having identified which ${\cal N}=1$ multiplets occur, the full spectrum 
can be deduced by diagonalizing the fermion mass terms. As for the gauge 
fields, this proceeds by expanding each adjoint fermion field as in 
(\ref{expan1}). It turns out that the extra chiral multiplets are degenerate 
with the massive vector multiplets and hence that {\em all}  
states in the theory have the masses given by (\ref{mass3}). For each value 
of $\vec{l}=(l_{1},l_{2})\neq \vec{0}$, 
we have a single massive vector multiplet and two massive chiral multiplets
\footnote{To be more precise, the mass formula 
for the different fermion species differ from (\ref{mass3}) 
by replacements of the form 
$(l_{1},l_{2}) \rightarrow (l_{1}\pm 1, l_{2}\pm 1)$ which leave the 
complete spectrum invariant.} 
with mass $M^{(\vec{l})}$, each transforming in the adjoint representation 
of $U(m)$. Thus the matter content at each mass level is exactly 
that of a massive BPS-saturated vector multiplet of ${\cal N}=4$ 
supersymmetry.      
\paragraph{}
The interpretation of the spectrum described above is 
clearest in the large-$n$ limit. Specifically, as $n\rightarrow\infty$, 
we have an infinite tower of states 
labelled by two arbitrary integers $l_{1}$ and $l_{2}$ 
(which we hold fixed in the limit). In this case we 
can make the replacement $\sin(l_{i}\pi/n)\simeq l_{i}\pi/n$ for $i=1,2$ 
and the mass formula (\ref{mass3}) becomes, 
\begin{equation}
M^{(\vec{l})}= \sqrt{l_{1}^{2}
\left(\frac{2\pi|\alpha_{1}|}{n}\right)^{2}+l_{2}^{2}
\left(\frac{2\pi|\alpha_{2}|}{n}\right)^{2}}
\label{kkspectrum}
\end{equation}        
with a degeneracy identical to that of a single massive BPS multiplet 
of ${\cal N}=4$ supersymmetry transforming in the adjoint 
representation of $U(m)$ at each non-zero mass level. 
Remarkably this spectrum 
is identical to that of a six-dimensional $U(m)$ gauge theory with 
non-chiral ${\cal N}=(1,1)$ supersymmetry compactified down to four 
dimensions with SUSY preserving 
boundary conditions on a rectangular torus of sides,     
\begin{eqnarray}
R_{1}=\frac{n}{2\pi|\alpha_{1}|} & \qquad{} \qquad{} & 
R_{2}=\frac{n}{2\pi|\alpha_{2}|} \nonumber \\
\label{torusr}
\end{eqnarray}
The integers $l_{1}$ and $l_{2}$ appearing in (\ref{kkspectrum}) 
correspond to the quantized momenta of each field along the compact 
dimensions. 
\paragraph{}
At energies far below the compactification scale, 
$E<<1/R_{1}$, $1/R_{2}$ the six-dimensional theory reduces
to a four-dimensional ${\cal N}=4$ theory with gauge group 
$U(m)$ and gauge coupling 
$G_{4}^{2}=4\pi^{2}R_{1}R_{2}/G_{6}^{2}$ where 
$G_{6}$ is the six dimensional gauge coupling. This agrees with the 
low energy theory on the Higgs branch ${\cal H}_{m}$ provided that we 
identify,   
\begin{equation}
G_{6}^{2}= \frac{g^{2}n}{|\alpha_{1}||\alpha_{2}|}
\label{sixd}
\end{equation} 
In addition, the conserved momenta of the six-dimensional theory 
along the compactified directions 
appear as central charges in the ${\cal N}=4$ supersymmetry 
algebra of the low-energy theory. Consequently, 
the massive KK modes form BPS multiplets of the low-energy 
${\cal N}=4$ supersymmetry in agreement with the spectrum of the 
$\beta$-deformed theory.  
\paragraph{}
Although we started with a theory with ${\cal N}=1$ supersymmetry in four 
dimensions, the spectrum we find matches that of a compactified 
six-dimensional theory with a much larger supersymmetry algebra. 
In fact, the recovery of sixteen supercharges as $n\rightarrow \infty$ is 
very natural as the deformation parameter $\beta=2\pi/n$ tends to zero 
in this limit and we formally recover the ${\cal N}=4$ supersymmetry 
of the undeformed theory. In this limit, the entire classical spectrum 
consists of BPS multiplets of ${\cal N}=4$ supersymmetry. As the masses of 
BPS states are not renormalized in an ${\cal N}=4$ theory, this suggests that 
the classical mass formula (\ref{mass3}) becomes exact in the large-$n$ 
limit. The issue of quantum corrections will be discussed in more detail in 
Section 5.3 below.   
\paragraph{}
Setting one of the momenta equal to zero (say $l_{2}$) 
the $n\rightarrow\infty$ spectrum 
has the standard form of a Kaluza-Klein tower: 
$M=l_{1}/R_{1}$ where $l_{1}$ can take any integer value. Returning to 
the case of finite-$n$ we have instead, 
\begin{equation}
M=\frac{n}{\pi R_{1}}\sin\left(\frac{l_{1}\pi}{n}\right)
\label{finiten}
\end{equation}
Now the quantized momentum, $l_{1}$, is only defined modulo $n$ and 
therefore naturally takes values in ${\bf Z}_{n}$ rather than ${\bf Z}$. This 
`compactification' of the momentum space is precisely what happens when a 
continuous spatial dimension is discretized. In fact the full mass formula 
(\ref{mass3}) is exactly the one you would get by replacing the continuous 
torus in spacetime with an $n\times n$ lattice with spacings, 
\begin{eqnarray}
\varepsilon_{1}=\frac{2\pi R_{1}}{n}=\frac{1}{|\alpha_{1}|} 
& \qquad{} \qquad{} & 
\varepsilon_{2}=\frac{2\pi R_{2}}{n} =\frac{1}{|\alpha_{2}|} \nonumber \\
\label{lattice}
\end{eqnarray}   
and replacing the kinetic terms with appropriate finite differences. 
\paragraph{}
In summary the full spectrum of the $\beta$-deformed theory 
on the Higgs branch ${\cal H}_{m}$ coincides with that of a 
six-dimensional $U(m)$ gauge theory with two compact discrete dimensions 
and four non-compact continuous dimensions. While the lattice theory preserves 
${\cal N}=1$ supersymmetry, the full classical spectrum of 
toroidally compactified 
six-dimensional ${\cal N}=(1,1)$ supersymmetric 
$U(m)$ Yang-Mills theory is recovered in a `continuum' limit, 
$n\rightarrow\infty$, $\varepsilon_{i}\rightarrow 0$. The fact that the 
spectra of the two theories coincide suggests the theories themselves may 
actually be the same. To make this convincing we must also compare the 
interactions of the two theories and, in particular, understand how 
six-dimensional gauge invariance can emerge from our 
four-dimensional theory. This is the main topic of the next subsection.     
 
\subsection{The Lattice Action}     
\paragraph{}
In our chosen vacua, 
the scalar vacuum expectation 
values can be written as $\langle \Phi_{i}\rangle=\alpha_{i}\Gamma_{i}$ for 
$i=1,2$ with, 
\begin{eqnarray}
\Gamma_{1} = I_{(m)}\otimes U_{(n)} & \qquad{} \qquad{} &
\Gamma_{2} = I_{(m)}\otimes V_{(n)} \nonumber \\
\label{vacm2}
\end{eqnarray} 
and $\langle \Phi_{3}\rangle=0$. 
\paragraph{}
Following 
\cite{AF1}, we expand the scalar fields $\Phi_{1}$ and $\Phi_{2}$ 
around their vacuum values as,  
\begin{eqnarray}
\Phi_{i} &= &\left(I_{(N)}+\varepsilon_{i}H_{i}\right)
U_{i}\langle \Phi_{i} \rangle  \nonumber \\
\label{fluct}
\end{eqnarray}
for $i=1,2$. The fluctuations are parametrized in terms of two 
$N\times N$ Hermitian matrices $H_{1}$, $H_{2}$ and two $N\times N$ unitary 
matrices $U_{1}$ and $U_{2}$. The (spontaneously-broken) $U(N)$ 
gauge symmetry of the $\beta$-deformed theory acts on these fields as, 
\begin{eqnarray}
H_{i}\rightarrow GH_{i}G^{\dagger} \qquad{} & & 
U_{i}\Gamma_{i}\rightarrow GU_{i}\Gamma_{i}G^{\dagger} \nonumber \\
\label{hutrans}
\end{eqnarray}
where $G\in U(N)$ and $i=1,2$.  
\paragraph{}
It is instructive to focus on the 
action of the unitary fluctuations $U_{i}$, setting the other 
fluctuating fields to zero. On substituting (\ref{fluct}) for 
$\Phi_{1}$ and $\Phi_{2}$ in the 
scalar potential of the $\beta$-deformed theory we obtain, 
\begin{equation}
V = \frac{|\alpha_{1}|^{2}|\alpha_{2}|^{2}}{4g^{2}} \, \sum_{i\neq j} 
\, {\rm Tr}_{N} \left[ U_{i}
\left(\Gamma_{i}U_{j}\Gamma_{i}^{\dagger}\right)
\left(\Gamma_{j}U_{i}^{\dagger}\Gamma_{j}^{\dagger}\right)
U^{\dagger}_{j}\right]    
\label{eguchi}
\end{equation}
\paragraph{}
The zero-dimensional matrix model with action (\ref{eguchi}) 
is known as the twisted 
Eguchi-Kawai model and its equivalence to a {\em non-commutative} 
lattice gauge theory is well known. We will now review this connection 
following the derivation of this result given in \cite{Sz} (see in particular 
Section 7.2 of this reference).   
\paragraph{} 
To understand how a two-dimensional lattice emerges we will consider the 
expansion (\ref{expan1}) of an arbitrary $N\times N$ matrix field $A(y)$ 
in terms of the basis elements $J^{(\vec{l})}$ with $m\times m$ coefficients 
$a^{(\vec{l})}(y)$. Here and in the following, $y_{\mu}$ denotes the 
coordinates of the four-dimensional spacetime. Our explicit 
calculation of the spectrum described above 
indicates that the $m\times m$ matrix fields $a^{(\vec{l})}(y)$ 
are mass eigenstates with eigenvalue $M^{(\vec{l})}$. 
In our Kaluza-Klein interpretation of the spectrum described, the vector 
$\vec{l}\in {\bf Z}^{2}_{n}$ is identified with the discrete momentum 
on the two 
dimensional torus. Hence the modes $a^{(\vec{l})}$ should be thought 
of as the Fourier modes of a six dimensional lattice field. 
\paragraph{}
We can make this correspondence  
explicit by defining a periodic $n\times n$ lattice ${\cal L}$ with points, 
\begin{equation}
\vec{x}=(x_{1},x_{2})=(n_{1}\varepsilon_{1},n_{2}\varepsilon_{2}) \, \in \, 
{\cal L}           
\label{lattice2}
\end{equation}
where the integers $n_{1}$ and $n_{2}$ are defined modulo $n$. 
The lattice spacings $\varepsilon_{1}$ and $\varepsilon_{2}$ are defined in 
(\ref{lattice}) above. For any $N\times N$ matrix field in four dimensions 
$A(y)$ we then introduce a corresponding $m\times m$ lattice field 
${\cal A}(\vec{x},y)$ defined by, 
\begin{eqnarray}
{\cal A}(\vec{x},y) \, \, & = &  \frac{1}{n^{2}}\, 
\sum_{\vec{l}\in {\bf Z}_{n}^{2}}\, a^{(\vec{l})}(y) 
\exp\left(i\frac{l_{1}x_{1}}{R_{1}}+ i\frac{l_{2}x_{2}}{R_{2}}\right) 
\nonumber \\ & = & \frac{1}{n^{2}}
\sum_{\vec{l}\in {\bf Z}_{n}^{2}}\, a^{(\vec{l})}(y) 
\exp\left(\frac{2\pi i}{n} \vec{l}\cdot \vec{n} \right) \nonumber \\ 
\label{latfield}
\end{eqnarray}
where the coefficients $a^{(\vec{l})}(y)$ are defined as in (\ref{expan1}) 
and $\vec{n}=(n_{1},n_{2})\in {\bf Z}_{n}^{2}$. By standard standard properties 
of the basis elements $J^{(\vec{l})}$ defined in (\ref{jl}), this 
defines a one-to-one invertible map 
$\Omega$ between $N\times N$ matrix fields in 
four dimensions and $m\times m$ matrix fields defined on the product 
of four-dimensional spacetime with the lattice ${\cal L}$ defined above. 
\begin{eqnarray}
\Omega\, :\,\,\, A(y) & \leftrightarrow & \Omega[A]={\cal A}(\vec{x},y) 
 \, \nonumber \\ 
\label{map}
\end{eqnarray}
The map is a finite dimensional version of the Wigner-Weyl 
correspondence between classical fields and operators. For simplicity 
we will suppress dependence on the four-dimensional coordinate $y_{\mu}$ 
in the following.   
\paragraph{}
Applying the map $\Omega$ to the unitary fluctuation 
matrices $U_{1}$ and $U_{2}$ yields two $m\times m$ lattice fields: 
${\cal U}_{i}(\vec{x})=\Omega [U_{i}]$ for $i=1,2$. 
Thus we repackage the $n^{2}m^{2}$ degrees of freedom 
in each $N\times N$ matrix $U_{i}$ as an 
$m\times m$ matrix field ${\cal U}_{i}(\vec{x})$ defined on an $n\times n$ 
lattice. The significance of this becomes clear when we rewrite the 
Eguchi-Kawai action (\ref{eguchi}) in terms of the lattice fields. 
The result is, 
\begin{equation}
V = \frac{|\alpha_{1}|^{2}|\alpha_{2}|^{2}}{4g^{2}n} \, \sum_{i\neq j}\, 
\sum_{\vec{x}\in {\cal L}}\, {\rm Tr}_{m}\left[
{\cal U}_{i}(\vec{x})
\star
{\cal U}_{j}(\vec{x}+\epsilon_{i}\hat{i})
\star
{\cal U}^{\dagger}_{i}(\vec{x}+\epsilon_{j}\hat{j})
\star 
{\cal U}^{\dagger}_{j}(\vec{x})\right]
\label{wilson}
\end{equation}       
where $\hat{i}$ denotes a unit vector in the $i$'th direction for $i=1,2$. 
We have also introduced a star-product of lattice fields which is the image 
of $N\times N$ matrix multiplication under the map $\Omega$ defined above. 
For any two lattice 
fields ${\cal A}(\vec{x})=\Omega[A]$ and ${\cal B}(\vec{x})=\Omega[B]$ 
we define ${\cal A}\star{\cal B}=\Omega[AB]$. Thus the  
unitary matrices $U_{i}$ are mapped to {\em star-unitary} lattice fields 
satisfying  ${\cal U}_{i}(\vec{x})\star {\cal U}^{\dagger}_{i}(\vec{x})= 
{\cal U}_{i}^{\dagger}(\vec{x})\star{\cal U}_{i}(\vec{x})=I_{(m)}$. 
As we now explain, (\ref{wilson}) is the action of a non-commutative 
lattice gauge theory. 
\paragraph{}
The explicit formula for the 
lattice star product is,   
\begin{equation} 
{\cal A}(\vec{x})\star {\cal B}(\vec{x})=\frac{1}{n^{2}}
\, \sum_{\vec{y},\vec{z}\in {\cal L}}\, 
{\cal A}(\vec{x}+\vec{y}) {\cal B}(\vec{x}+\vec{z})\, \exp\left(2i\, 
\vec{y}\cdot \vartheta^{-1}\cdot \vec{z}\right)
\label{starprod}
\end{equation}
where the two-form $\vartheta_{ij}=\vartheta\varepsilon_{ij}$ with, 
\begin{equation}
\vartheta =\frac{n}{2\pi}
\varepsilon_{1}\varepsilon_{2}
\label{noncom1}
\end{equation}
The lattice star-product is a discretized 
form of the Groenewald-Moyal (GM) star product used to define quantum 
field theory on a continuum non-commutative space. In particular, 
the lattice sums appearing in (\ref{starprod}) are finite-dimensional 
approximations to the space-time integrals appearing in the standard 
definition of the continuum GM star-product. 
Like its continuum counterpart, the lattice star-product 
is associative but non-commutative even in the abelian case $m=1$. 
The length-scale at which 
non-commutativity becomes significant is set by the parameter 
$\sqrt{\vartheta}$. Taking a formal limit in which $\vartheta\rightarrow 0$ 
the lattice star product becomes an ordinary product of $m\times m$    
matrices. In this limit, the lattice action 
(\ref{wilson}) becomes precisely the standard Wilson placquette action for 
a $U(m)$ lattice gauge theory. The action (\ref{wilson}) can therefore be 
thought of as a non-commutative deformation of $U(m)$ lattice gauge theory
and it was proposed in \cite{AMNS} (see also \cite{Nish}) as a proper non-perturbative definition 
of the corresponding continuum non-commutative gauge theory.    
\paragraph{} 
For non-zero $\vartheta$, the $U(m)$ gauge symmetry of the Wilson action is 
deformed to a star-gauge symmetry. In the present context 
this symmetry is actually a direct consequence of the $U(N)$ gauge symmetry 
of the $\beta$-deformed theory. For each element $G\in U(N)$, 
we introduce an $m\times m$ matrix 
field ${\cal G}(\vec{x})=\Omega[G]$ defined at each lattice point.  
As $G$ is unitary, its image ${\cal G}$ under $\Omega$ is constrained to be 
star-unitary: ${\cal G}(\vec{x}) \star{\cal G}^{\dagger}(\vec{x})= 
{\cal G}^{\dagger}(\vec{x})\star {\cal G}(\vec{x})=I_{(m)}$. The invariance 
of the $\beta$-deformed theory under the $U(N)$ gauge transformation 
(\ref{hutrans}) then implies that the  
non-commutative lattice action (\ref{wilson}) is then invariant under the 
star gauge transformation;  
\begin{eqnarray}
{\cal U}_{i}(\vec{x}) &\, \, \rightarrow\, \, 
& {\cal G}(\vec{x})\star{\cal U}_{i}(\vec{x})\star
{\cal G}^{\dagger}(\vec{x}+\varepsilon_{i}\hat{i}) \nonumber \\
\label{stargauge}
\end{eqnarray}
For $\vartheta\rightarrow 0$ this reduces to the standard $U(m)$ gauge 
invariance of the Wilson lattice action. 
\paragraph{}
So far we have only considered the unitary fluctuations $U_{1}$ and $U_{2}$ 
of the scalar fields $\Phi_{1}$ and $\Phi_{2}$ which give rise to 
lattice gauge fields ${\cal U}_{i}(\vec{x})$. Among the other fields we 
need to restore are the Hermitian fluctuations $H_{1}(y)$ and $H_{2}(y)$ 
defined in (\ref{fluct}). Under the map $\Omega$, these yield $m\times m$ 
lattice fields ${\cal H}_{i}=\Omega[H_{i}]$ for $i=1,2$. 
As in (\ref{hutrans}), the fields $H_{i}$ transform in the adjoint 
representation of the $U(N)$ gauge group. This means that the resulting 
lattice fields ${\cal H}_{i}$ transform in the adjoint of the 
$U(m)$ star-gauge transformation,   
\begin{eqnarray}
{\cal H}(\vec{x}) &\, \, \rightarrow\, \, 
& {\cal G}(\vec{x})\star{\cal H}(\vec{x})\star
{\cal G}^{\dagger}(\vec{x}) \nonumber \\
\label{stargauge2}
\end{eqnarray}
for $i=1,2$. 
Note the difference between the transformation properties of $H_{i}$ and 
$U_{i}$ in (\ref{hutrans}) which leads to the 
different transformations (\ref{stargauge2}) and (\ref{stargauge}) for the 
corresponding lattice fields. Thus the fields ${\cal H}_{i}$ transform 
like adjoint-valued scalar fields in the resulting 
six-dimensional theory, while the ${\cal U}_{i}$ correspond to 
the components of a six-dimensional gauge field in the two discrete 
directions.    
\paragraph{}
By applying the map $\Omega$ to each field of the $\beta$-deformed theory, 
it is possible to rewrite the whole action of the four-dimensional 
theory on the Higgs branch ${\cal H}_{m}$ as a six-dimensional 
non-commutative $U(m)$ gauge theory with two compact discrete dimensions 
\cite{AF1}. The resulting action is invariant under $U(m)$ 
star-gauge transformations and under 
Lorentz transformations of the four continuous dimensions. It also invariant 
under the action of four supercharges inherited from the 
$\beta$-deformed theory. In fact the resulting theory is a 
discretized version of six-dimensional non-commutative 
${\cal N}=(1,1)$ supersymmetric Yang-Mills theory with gauge group 
$U(m)$. This is consistent with the calculation of the spectrum 
reported in the previous subsection. 
We will not give the full lattice action here, 
briefly review the origin of the six-dimensional fields and their 
interactions.    
\paragraph{}
Under $\Omega$, 
the four-dimensional $U(N)$ gauge field yields the components of a 
$U(m)$ gauge field along the four continuous dimensions. Together with 
the lattice gauge fields ${\cal U}_{i}$, $i=1,2$, they define the 
six dimensional 
$U(m)$ gauge field. The remaining complex scalar field $\Phi_{3}$ yields a 
complex adjoint scalar in six dimensions. Together with the fields  
${\cal H}_{i}$ $i=1,2$ we have a total of four real adjoint scalars. 
The fermionic content of the theory includes four 
species of complex adjoint fermions which transform as Weyl fermions under 
Lorentz transformations of the four continuous dimensions.
This is precisely the matter content of a $U(m)$ vector multiplet of 
${\cal N}=(1,1)$ supersymmetry in six dimensions.         
Each of these fields has a discretized kinetic term and the spacetime 
derivatives appearing are covariant with respect to star-gauge 
transformations. The action also includes a commutator potential for the 
adjoint scalars which matches that of the ${\cal N}=(1,1)$ theory 
in six dimensions. The interactions of the fermionic fields 
are determined by the four unbroken supersymmetries of the 
lattice action.             
\paragraph{}
The non-commutative lattice action (\ref{wilson}) has a standard classical 
continuum limit obtained by taking an $n\rightarrow \infty$ limit in which the 
lattice spacings $\varepsilon_{i}$ go to zero while holding the length-scale 
of non-commutativity 
$\sqrt{\vartheta}=\sqrt{n\varepsilon_{1}\varepsilon_{2}/2\pi}$ fixed. 
The resulting scaling $\varepsilon_{i}\sim 1/\sqrt{n}$ for $i=1,2$, mean 
that the radii of the torus $R_{i}=n\varepsilon_{i}/2\pi$ go to infinity 
like $\sqrt{n}$. In this limit the lattice action (\ref{wilson}) goes over 
to the Maxwell action of a $U(m)$ gauge theory on an infinite 
non-commutative plane $R^{2}_{\vartheta}$. 
Restoring the extra four commutative dimensions of the 
$\beta$-deformed theory and the additional field content described above 
we recover a maximally supersymmetric six-dimensional $U(m)$ gauge theory on 
$R^{3,1}\times R^{2}_{\vartheta}$.    
\paragraph{}    
As explained in \cite{AF1}, we may also understand this result by making 
contact with the ideas of Matrix theory \cite{BFSS}. As before, we 
expand the fields of the $U(N)$ 
$\beta$-deformed theory around their background 
values $\langle \Phi_{i}\rangle =\alpha_{i}\Gamma_{i}$ and take the 
$n\rightarrow \infty$ limit. As the deformation parameter $\beta=2\pi/n$ 
goes to zero in this limit we can simply perform the calculation using the 
undeformed ${\cal N}=4$ Lagrangian. For $i=1,2$, 
we formally introduce new fields, 
\begin{equation}
\Phi_{i}=\exp\left(i\varepsilon_{i}\hat{C}_{i}\right)\simeq I_{(N)}+
i\varepsilon_{i}\hat{C}_{i}+ \,\, O(\varepsilon^{2}) 
\label{formal}
\end{equation}         
where the second equality holds near the continuum limit 
$n\rightarrow \infty$, $\varepsilon_{i}\sim 1/\sqrt{n}\rightarrow 0$. 
Using the second equality (and performing a suitable rescaling) 
we can effectively replace $\Phi_{i}$ by $\hat{X}_{i}=\vartheta_{ij}\hat{C}_{j}$ 
in the ${\cal N}=4$ action where the non-commutativity tensor 
$\vartheta_{ij}$ is defined above. 
The matrices $\Gamma_{i}$ appearing in the vacuum values of the 
$\Phi_{i}$ satisfy $\Gamma_{1}\Gamma_{2}=
\bar{\omega}_{n}\Gamma_{2}\Gamma_{1}$. As a consequence, in the vacuum state, 
the fields $\hat{X}_{i}$ satisfy,
\begin{equation}
[\hat{X}_{i},\hat{X}_{j}]=i\vartheta_{ij}
\label{ncplane}   
\end{equation}
This relation has no solutions for finite $N=mn$, and the required background 
only exists for $N=\infty$. 
Expanding the fields of a maximally supersymmetric $U(\infty)$ 
gauge theory in $p$ 
spacetime dimensions, around this background yields a maximally 
supersymmetric, non-commutative $U(m)$ gauge theory in $p+2$ dimensions 
\cite{Mat,Seib}.   
Indeed, this is just the standard Matrix theory construction of the 
worldvolume theory on $m$ D(p+2) branes starting from the theory on 
$N=mn$ Dp branes. 
\paragraph{}
Another interesting connection discussed in \cite{AF1} 
is to the original notion of deconstruction introduced in 
\cite{AHCG,AHCK}. In particular, a four-dimensional gauge theory 
which deconstructs the {\em commutative} version of the 
six-dimensional ${\cal N}=(1,1)$ 
theory considered here was given in \cite{AHCK}. To deconstruct the 
six-dimensional theory with gauge group $U(m)$ one needs a four-dimensional 
${\cal N}=1$ quiver theory with gauge group $U(m)^{n}$ and certain 
bifundamental chiral multiplets. From a field theoretic perspective there 
is no obvious connection between this theory and the 
$\beta$-deformed theory with gauge group $U(N)$. 
\paragraph{}
As explained in \cite{AF1}, 
the connection only becomes clear when one considers the string theory 
realisation of the two theories. The quiver theory of \cite{AHCK} is the 
world volume theory of $m$ D3 branes placed at a 
${\bf C}^{3}/{\bf Z}_{n}\times {\bf Z}_{n}$ singularity. As we mentioned in 
Section 3, the $\beta$-deformed theory on ${\cal H}_{m}$ lives on the 
world-volume of $m$ D3 branes at the same singularity but now with a single 
unit of discrete torsion. A string-theoretic argument given in 
\cite{AHCK} explains why the orbifold theory naturally gives rise to a 
six-dimensional lattice gauge theory. One of the main points of \cite{AF1} 
is that the same argument goes through in the presence of discrete 
torsion. The sole modification caused by the introduction of discrete torsion 
is that the resulting lattice gauge theory becomes 
non-commutative.                  
\paragraph{}
Finally we could also repeat the analysis of this Section in a more general 
Higgs branch vacuum with ${\cal H}_{m}$ with $\alpha_{3}\neq 0$. It is 
claimed in \cite{AF1} that this leads to a non-rectangular or slanted 
torus. We have not checked this directly, but it will be confirmed in the 
string theory construction of Section 8.       
\section{The Low-Energy Effective Theory}
\subsection{The Classical Theory}
In the previous Section, we have argued that, for $N=nm$, the 
$U(N)$ $\beta$-deformed theory in the Higgs branch vacuum (\ref{vacm})
specified by complex numbers $\alpha_{1}$ and $\alpha_{2}$ 
is classically equivalent to a certain six-dimensional theory. The 
six-dimensional theory is a $U(m)$ gauge theory compactified on an 
$n\times n$ lattice with periodic boundary conditions. Collecting 
various formulae given above, the parameters of the six dimensional theory 
can be expressed in terms of the four-dimensional parameters. 
The lattice spacings of the two compact discrete dimensions are, 
\begin{eqnarray}
\varepsilon_{1}=\frac{1}{|\alpha_{1}|} & \qquad{} \qquad{} \qquad 
& \varepsilon_{2}=\frac{1}{|\alpha_{2}|} \nonumber \\
\label{table1}
\end{eqnarray}
The radii of the two-dimensional torus which they define are, 
\begin{eqnarray}
R_{1}=\frac{n}{2\pi |\alpha_{1}|} & \qquad{} \qquad{} \qquad{} 
& R_{2}=\frac{n}{2\pi|\alpha_{2}|} \nonumber \\
\label{table2}
\end{eqnarray}
The six-dimensional gauge coupling and non-commutivity parameter are given 
by, 
\begin{eqnarray}
G_{6}=\frac{g^{2}n}{|\alpha_{1}||\alpha_{2}|} & \qquad{} \qquad{} \qquad{} & 
\vartheta=\frac{n}{2\pi|\alpha_{1}||\alpha_{2}|}  \nonumber \\
\label{table3}
\end{eqnarray}
respectively. As is conventional, we will also define a 
dimensionless non-commutativity parameter $\Theta=\vartheta/2\pi R_{1}R_{2}=1/n$. 
\paragraph{}
Importantly our derivation of the six dimensional action 
was based on a classical analysis of the ${\beta}$-deformed theory. 
This is only valid when the four dimensional t' Hooft couping 
$\lambda=g^{2}N$ is much less than one. In this regime the dynamical-length 
scale of the six-dimensional gauge theory set by the gauge coupling 
$G_{6}$ is much smaller than the lattice spacing $\varepsilon$. Thus the 
six-dimensional lattice gauge theory is far 
from its continuum limit. However, even at weak coupling, 
we can obtain an effective continuum theory by focussing 
on modes of wavelength much larger than the 
lattice spacing. Writing a classical effective action for these modes 
we can replace discrete derivatives by continuous ones and obtain a 
continuum effective action. This proceedure is identical to taking the 
classical continuum limit discussed above. Hence our long-wavelength 
effective action, valid on length scales much larger than 
${\rm max}\{\varepsilon_{1},\varepsilon_{2}\}$, is ${\cal N}=(1,1)$ 
supersymmetric $U(m)$ gauge theory on $R^{3,1}\times T^{2}_{\Theta}$. 
\subsection{Morita Equivalence}
\paragraph{}
As the the effective theory is defined on a non-commutative torus with 
rational non-commutativity parameter, $\Theta=1/n$,  
it can be related by Morita duality 
to a commutative theory (See \cite{Sz} for a review of this topic). 
In the present case, the Morita dual theory is a 
six dimensional ${\cal N}=(1,1)$ gauge theory with gauge group 
$G=U(N)$ defined on the commutative space $T^{2}\times R^{3,1}$. The 
commutative torus in question is rectangular and has sides has 
sides of length; 
\begin{eqnarray}
R'_{1}=\frac{1}{n}R_{1}=\frac{1}{2\pi|\alpha_{1}|} & \qquad{} 
\qquad{} &   R'_{2}=\frac{1}{n}R_{2}=\frac{1}{2\pi|\alpha_{2}|} 
\nonumber \\
\label{Isides}
\end{eqnarray}
The dual six-dimensional gauge coupling is, 
\begin{equation}
G_{6}^{'2}=\frac{{G}^{2}_{6}}{n}= 
\frac{g^{2}}{|\alpha_{1}||\alpha_{2}|}
\label{Icoupling}
\end{equation}
The theory has non-trivial boundary conditions on $T^{2}$. If we choose 
coordinates $\vec{x}=(x_{1},x_{2})$ on $T^{2}$ with 
$0\leq x_{1}< 2\pi R'_{1}$ and $0\leq x_{2}< 2\pi R'_{2}$, then all 
adjoint fields obey the 't Hooft boundary conditions; 
\begin{eqnarray}
A'(x_{1}+2\pi R'_{1},x_{2}) & = 
& \Gamma_{1}A'(x_{1},x_{2})\Gamma_{1}^{\dagger} 
\nonumber \\
A'(x_{1}, x_{2}+2\pi R'_{2}) & = 
& \Gamma_{2}A'(x_{1},x_{2})\Gamma^{\dagger}_{2} 
\nonumber \\
\label{thooft}
\end{eqnarray}    
where the matrices $\Gamma_{1}$ and $\Gamma_{2}$ are defined in (\ref{vacm2}) 
above. 
\paragraph{}
The equivalence is a one-to-one map between classical 
field configurations in the two 
theories. We start from the non-commutative $U(m)$ effective theory, 
choosing coordinates $\hat{\vec{x}}=(\hat{x}_{1},\hat{x}_{2})$ 
on the torus $T^{2}_{\Theta}$ with $0\leq \hat{x}_{1}< 2\pi R_{1}$ and 
$0\leq \hat{x}_{2}< 2\pi R_{2}$. Every single valued 
adjoint field on $T^{2}_{\Theta}$ has a 
Fourier expansion of the form, 
\begin{eqnarray}
{\cal A}(\vec{x},y) & = &\frac{1}{n^{2}}\sum_{\vec{l}\in {\bf Z}^{2}}\, 
a^{(\vec{l})}(y)\, \exp\left(2\pi i\left[
\frac{l_{1}\hat{x}_{1}}{R_{1}}+\frac{l_{2}\hat{x}_{2}}{R_{2}}\right]\right) \nonumber \\
\label{fourierII}
\end{eqnarray}
where, as above, $y_{\mu}$ denotes the coordinates of the four commutative 
directions. This expansion is just the continuum version of the expansion 
(\ref{latfield}) of the corresponding field defined on the lattice ${\cal L}$. 
In particular, the coefficients $a^{(\vec{l})}$, 
with $\vec{l}\in {\bf Z}^{2}$, appearing in 
(\ref{fourierII}) are the large-$n$ limit of the corresponding coefficients 
in (\ref{latfield}) where $\vec{l}\in {\bf Z}^{2}_{n}$.  
\paragraph{}
A choice of the $m\times m$ Fourier modes $a^{\vec{l}}(y)$ 
of each adjoint-valued 
field uniquely defines a field configuration in our $U(m)$ effective theory. 
We can define a 
corresponding field configuration in the dual $U(N)$ theory, by using the same 
Fourier coefficients. Specifically each adjoint-valued field of the dual 
theory is an $N\times N$ matrix which we write as,
\begin{eqnarray}
{\cal A}'(\vec{x},y) & = & \frac{1}{n^{2}} \sum_{\vec{l}\in {\bf Z}^{2}}\,
\, a^{(\vec{l})}(y)\,\otimes \,
V^{l_{1}}_{(n)}U_{(n)}^{-l_{2}}\omega^{\frac{l_{1}l_{2}}{2}}_{n} 
\exp\left(2\pi i\left[\frac{l_{1}x_{1}}{nR'_{1}}+\frac{l_{2}x_{2}}{nR'_{2}}
\right]\right) \nonumber \\
\label{fourierI}
\end{eqnarray}        
It is straightforward to check that this field configuration 
satisfies the approriate 
't Hooft boundary conditions (\ref{thooft}) on the dual commutative torus.    
\paragraph{}
The precise statement of classical 
Morita equivalence is that the Yang-Mills action 
of the configurations ${\cal A}$ and ${\cal A}'$ 
is the same. As will be important in the following, 
various topological charges carried 
by the two dual field configurations are mapped onto each other \cite{Mor}. 
As in lower 
dimensions the equivalence between the two theories can be thought of as a 
formal change of variables in the path integral. In dimension less than or 
equal to four this implies an exact equivalence between the theories at the 
quantum level which can actually be realised graph by graph in the 
Feynman diagram expansion. In the present case, both continuum theories are 
non-renormalisable and so the status of Morita equivalence in the quantum 
theory is less clear. In the following we, will only use the equivalence as a 
map between classical configurations. 
\paragraph{}
For the present purposes, the key point is that the dual six-dimensional 
$U(N)$ formulation 
of the low-energy effective theory can actually be related to the original 
four-dimensional $U(N)$ gauge theory we started with. To make contact 
with the four dimensional theory we simply dimensionally reduce the 
six-dimensional dual gauge theory to a four dimensional one by setting 
$\vec{x}=\vec{0}$ in (\ref{fourierI}). Comparing (\ref{fourierII}) with 
(\ref{latfield}) and (\ref{fourierI}) with (\ref{expan1}), 
we see that the relation between the six dimensional $U(m)$ adjoint field 
$A(\vec{x},y)$ and the four-dimensional $U(N)$ adjoint field 
$A'(\vec{0},y)$ is simply the large-$n$ limit of the the one-to-one map 
$\Omega$ between matrices and 
lattice fields introduced in the 
previous section. The upshot of this is that, at large-$n$, we have an 
explicit map between the six-dimensional $U(m)$ fields of the 
effective theory and the $U(N)$ fields of the four-dimensional 
$\beta$-deformed theory we started with. 
 For example, if we take the components 
${\cal A}_{\mu}(\vec{x},y)$ for $\mu=0,1,2,3$ of the $U(m)$ gauge field in the 
four commutative directions, we can actually identify 
the dual $U(N)$ field ${\cal A}'_{\mu}(\vec{0},y)$ with the $U(N)$ gauge 
field of the four-dimensional $\beta$-deformed gauge theory. In the 
next section, we will use this correspondence to interpret the topological 
charges of the $U(m)$ effective theory in terms of those of 
the four-dimensional theory.   
\subsection{Quantum Corrections}                 
\paragraph{}
The classical derivation of the effective theory presented above is only 
valid at weak 't Hooft coupling: $g^{2}N<<1$. In this subsection , we 
will consider how the classical picture is modified by quantum corrections.    
As explained in Section 3, the Higgs branch ${\cal H}_{m}$ cannot be lifted 
because of the ${\cal N}=1$ superconformal invariance and the other 
non-anomalous global symmetries. Thus the massless spectrum will remain the 
same in the quantum theory. In Section 3, we argued that the complex 
structure of the Higgs branch is also protected from quantum corrections 
by the symmetries of the theory.   
\paragraph{}
On the other hand, it is 
obvious that the low-energy effective theory with parameters as given in 
(\ref{table1},\ref{table2},\ref{table3}) must be modified in the 
quantum theory. Gauge theories in six dimensions become strongly-coupled 
in the UV. In particular, we expect that perturbation theory breaks down on 
length-scales smaller than the scale, $\sqrt{G_{6}^{2}m}$, set by the 
six-dimensional 't Hooft coupling. Thus the low-energy theory described above 
only makes sense when this scale is much smaller than the lattice spacing 
$\varepsilon$. From (\ref{table1}) and (\ref{table3}), we see that 
this is only true when $g^{2}N<<1$. Away from weak coupling, the 
situation is less clear. If the lattice spacing $\varepsilon$ is really 
smaller than the lengthscale set by the 't Hooft coupling, 
then the effective theory cannot simply be a six dimensional gauge theory: 
some new physics must be included on length-scales of order 
$\sqrt{G_{6}^{2}m}=\sqrt{g^{2}N/|\alpha_{1}||\alpha_{2}|}$. We will start to 
address this issue in Sections 8 and 9 and will consider it further 
in \cite{nd2}. 
\paragraph{}
The theory on the Higgs branch ${\cal H}_{m}$, 
exhibits a key simplification in the 
large-$N$ limit which we have already touched on above. As the 
deformation parameter $\beta=2\pi/n$, with $n=N/m$, the theory recovers 
${\cal N}=4$ supersymmetry in a large-$N$ limit with $m$ and $g^{2}$ 
fixed. Note that the additional supersymmetry 
appears at the level of the microscopic action and is 
therefore valid on all length-scales and for any fixed value of the 
gauge coupling.  In this limit the four-dimensional 't Hooft coupling becomes 
large, $g^{2}N>>1$ and perturbation theory breaks down. 
Instead the UV theory is well 
described by IIB supergravity (provided $g^{2}<<1$). 
In addition to taking a large-$N$ limit, we are simultaneously 
taking the deformation parameter $\beta=2\pi m/N$ to zero. 
The fact that the $\beta$-deformation has a smooth 
supergravity description suggests that this 
combined limit exists and makes sense. 
\paragraph{}
The spectrum discussed in Section 
4.1, consists entirely of BPS-saturated multiplets of the enlarged 
${\cal N}=4$ supersymmetry. Their masses are determined by the 
central charges of the ${\cal N}=4$ supersymmetry algebra which 
correspond to momenta in the two compact dimensions. In theories with 
${\cal N}=4$ supersymmetry, central charges do not recieve quantum 
corrections. For this reason, we expect that the classical mass formula 
(\ref{mass3}) becomes exact in the $N\rightarrow \infty$ limit described 
above. As the enlarged supersymmetry is valid for all length scales, this 
should be true even for states near the edge of the Brillouin zone 
with momenta $l_{i}\sim n=N/m$. The enlarged supersymmetry also forces the 
effective action for the massless degrees of freedom to be 
${\cal N}=4$ SUSY Yang-Mills with gauge group $U(m)$ as it was in the 
classical case. The low energy gauge coupling is exactly given by the 
replacement $\tau\rightarrow \tau_{R}$ in the classical formula 
$\tau_{H}=N\tau$.   
\paragraph{}         
To summarize our conclusions, the Higgs branch ${\cal H}_{m}$ exists and 
has the same massless spectrum for all 
values of the parameters. Away from the semiclassical regime 
$g^{2}N<<1$, the effective theory cannot simply be the six-dimensional 
gauge theory which appears classically. However, in the 
$N\rightarrow \infty$ limit with fixed $g^{2}$, the spectrum of 
Kaluza-Klein modes and the effective action for the massless degrees of 
freedom are exactly the same as they are in the classical theory.        

\section{Instanton Strings and Magnetic Confinement}
\paragraph{}                     
In this section we will focus on the classical solutions of the $U(m)$ 
effective theory. 
An important fact about non-abelian gauge theories 
in (5+1)-dimensions is that they possess exact classical solutions 
corresponding to finite tension strings. The solutions are obtained by 
embedding the familiar Yang-Mills instantons in four of the five spatial 
dimensions of the effective theory. The resulting object is then 
interpreted as a string infinitely stretched in the 
remaining spatial dimension. The string tension 
can be deduced from the action of a 4d instanton; 
\begin{equation}
T=\frac{8\pi^{2}}{G^{2}_{6}}=
\frac{8\pi^{2}|\alpha_{1}||\alpha_{2}|}{g^{2}n}
\label{tension}
\end{equation}
The string tension $T$ saturates a Bogomol'nyi lower bound and the 
corresponding self-dual 
gauge field configuration preserves eight of the sixteen 
supercharges of the ${\cal N}=(1,1)$ theory. 
We will refer to these objects as BPS strings in the following.
It is important to bear in mind that these strings are BPS solutions only 
of the low-energy effective theory. At finite $N$, they are certainly not 
BPS configurations of the full theory which only has four supercharges 
and has no central charges corresponding to strings or particles. As 
$N\rightarrow \infty$ the sixteen unbroken supersymmetries are recovered. 
Similar arguments to those given in the previous section suggest that the 
semiclassical formula (\ref{tension}) for the tension 
becomes exact in this limit.           
\paragraph{}
The classical effective action introduced above 
can safely be used to describe 
field configurations whose typical spatial variation in the two 
discretized dimensions has wavelength much larger than the 
lattice spacing. The gauge fields of ordinary non-abelian 
Yang-Mills instanton are 
characterised by a variable scale size $\rho$ and they will be correctly 
described by the continuum effective action if this size is much larger 
than the lattice spacing; $\rho>>{\rm max}
\{\varepsilon_{1},\varepsilon_{2}\}$.
\paragraph{}       
As the effective theory is formulated on 
$T^{2}_{\Theta}\times R^{3,1}$ (or 
$T^{2}_{\Theta}\times R^{4}$ in the Euclidean case), the Yang-Mills instanton 
gives rise to three basic types of configuration: 
\paragraph{}
{\bf Long Strings} In this case we consider a static string of infinite 
length stretched in one of the three spatial directions of $R^{3,1}$. The 
classical field configuration is equivalent to a single Yang-Mills 
instanton of a four-dimensional Euclidean $U(m)$ gauge theory on the 
non-commutative space $T^{2}_{\Theta}\times R^{2}$ with $\Theta=1/n$. 
\paragraph{}
{\bf Winding Modes} Here we consider a static string wound $q_{1}$, $q_{2}$  
times around the $x_{1}$, $x_{2}$ dimensions of 
$T^{2}_{\Theta}$. This gives rise to a towers of 
BPS solitons with masses, 
\begin{equation}
M^{2}_{(q_{1},q_{2})}= T^{2}\left(q_{1}^{2}R_{1}^{2}+
q_{2}^{2}R_{2}^{2}\right)=
\left(\frac{8\pi^{2}}{g^{2}}\right)^{2}
\left(q_{1}^{2}|\alpha_{2}|^{2}+q_{2}^{2}|\alpha_{1}|^{2}\right)
\label{winding}
\end{equation}
Configurations with $q_{2}=0$ correspond to $q_{1}$ self-dual $U(m)$ 
Yang-Mills instantons on $R^{3}\times S^{1}$ where the radius of the compact 
dimension is $R_{1}=n/2\pi|\alpha_{1}|$.     
\paragraph{}
{\bf Worldsheet Instantons} Finally, in the Euclidean theory, 
we can consider the case of a Euclidean string worldsheet wrapped $p$ 
times around the torus $T^{2}_{\Theta}$. This yields an instanton of 
action,   
\begin{equation}
S=pT\, 4\pi^{2}R_{1}R_{2}= \frac{8\pi^{2}np}{g^{2}}
\label{inst}
\end{equation}
\paragraph{} 
Each of the objects listed above can be realised as a classical solution 
of the six-dimensional $U(m)$ effective theory. As discussed above, the 
description will be valid provided the scale-size of the corresponding 
instanton is much 
larger than the lattice spacing. The formulae of the previous section 
allow us to map these back to configurations of the original $U(N)$ fields 
of the $\beta$-deformed theory. In the rest of this section we will see that 
the long strings discussed above actually correspond to stable classical 
solutions of the full $\beta$-deformed theory, labelled by a 
conserved topological charge. More generally, some of the other 
configurations discussed above may not correspond solutions of the 
full theory and may have instabilities. 
In this case, the prediction of deconstruction is that 
these objects become stable in the large-$n$ limit.  
In the quantum theory at finite $n$, these will correspond to unstable states 
with typical life time $\tau$ proportional to a positive power of $n$.     
\paragraph{}        
In a Higgs phase vacuum, 
we typically expect to find the confinement of magnetic charges. 
The usual signal of this confinement is the formation of 
magnetic flux tubes of finite tension which can end on magnetic charges. 
Now we have seen that the low energy effective theory on 
${\cal H}_{m}$ does indeed possess string-like solutions and it is natural to 
guess that these strings are precisely the expected flux tubes. 
In the following we will confirm this guess using both field 
theory and string theory arguments. 
\paragraph{}
The relation between the Higgs mechanism and magnetic 
confinement is clearest when the theory has a mass gap. In this case 
't Hooft's standard classification of possible massive phases is applicable 
\cite{TH}. 
However, for the $\beta$-deformed theory, there can never be a mass gap. 
In addition to the massless photon of 
the unbroken central $U(1)$ subgroup of $U(N)$, the Higgs-phase 
theory always has massless scalars due to the spontaneous breaking of 
conformal invariance. In theories with massless scalars, the existence 
of stable magnetic flux tubes is not guarenteed. Recent work by several 
authors \cite{Vortex} has illustrated several subtleties of the 
massless case and thrown 
doubt onto the occurence of magnetic confinement in these 
theories. In the present case we will see that, despite the presence of 
massless scalars, stable magnetic flux tubes do 
emerge in the large-$n$ limit.  
\paragraph{}
In any non-abelian gauge theory the possible magnetic charges and fluxes are 
classified by the first homotopy class of the gauge group. In the 
present case we have $\pi_{1}(U(N))={\bf Z}$. 
The non-abelian magnetic flux is therefore specified by the choice 
of an integer $k$. In a theory with all 
fields in the adjoint representation, the photon of the 
central $U(1)$ subgroup of $U(N)$ decouples\footnote{We remind the reader 
that this does not mean that the $U(N)$ and $SU(N)$ theories are related in a 
simple way. In particular the former theory has three 
additional neutral chiral multiplets.}.
Hence we decompose $U(N)$ in the standard way as, 
\begin{equation}
U(N)\equiv \frac{U(1)\times SU(N)}{{\bf Z}_{N}}
\label{u1}
\end{equation}
Under this decomposition, we can identify a $U(1)$ magnetic flux, $l\in 
\pi_{1}[U(1)]={\bf Z}$ (which is equal to the total $U(N)$ magnetic flux $k$) 
and the so-called 
't Hooft non-abelian magnetic flux 
$[m]\in \pi_{1}[SU(N)/{\bf Z}_{N}]={\bf Z}_{N}$. The decomposition (\ref{u1}) implies 
the selection rule $m=-l=-k$ mod $N$ which correlates the 
't Hooft flux with the total $U(N)$ flux. 
A flux tube carrying $k$ units of flux can end on a 
magnetic charge of quantized value $\pm k$  
The formation of stable finite-tension strings carrying integer multiples of 
magnetic flux $k$ leads to the linear confinement of objects 
carrying magnetic charges which are integer multiples of $\pm k$.              
\paragraph{}  
In the light of the above discussion we would like to show 
that the instanton string solutions of the six-dimensional 
low-energy effective theory 
actually correspond to configurations carrying $U(N)$ magnetic flux 
in the original $\beta$-deformed theory. This follows from the 
standard properties of the Morita equivalence discussed in the previous 
section. In particular we start from a long $U(m)$ instanton string, 
stretched in the $y_{3}$ direction. The string corresponds to a 
static classical configuration of the six-dimensional $U(m)$ gauge 
connection\footnote{In this section we will avoid 
introducing six-dimensional Lorentz indices and work in terms of 
differential forms instead} ${\cal A}(\vec{x},y)$.   
which depends only on the four spatial coordinates 
$(\hat{x}_{1},\hat{x}_{2}, y_{1},y_{2})$ which parameterize a submanifold 
$T^{2}_{\Theta}\times R^{2}_{12}$ of the full six-dimensional spacetime 
$T^{2}_{\Theta}\times R^{3,1}$. The instanton string is characterized by 
a non-zero second Chern class, 
\begin{equation}
k=\frac{1}{8\pi^{2}}\int_{T^{2}_{\Theta}\times R^{2}_{12}}\,\, {\rm Tr}_{m}
\left[{\cal F}\wedge {\cal F}\right]\, \in\, {\bf Z}
\label{2nd}
\end{equation}
where ${\cal F}$ is the corresponding curvature two-form.     
The Morita equivalence discussed in the previous section maps 
the non-commutative $U(m)$ connection ${\cal A}$ to a commutative  
$U(N)$ connection ${\cal A}'$ defined on $T^{2}\times R^{3,1}$ with '
t Hooft boundary conditions 
on $T^{2}$. By construction the resulting 
field configuration only depends on the four coordinates 
$(x_{1},x_{2},y_{1},y_{2})$. A standard property of the Morita map 
(see for example \cite{Mor}) is that it maps second Chern class 
(\ref{2nd}) in Theory II to a first Chern class of Theory I, defined as 
\begin{equation}
k'=\frac{1}{2\pi}\int_{R^{2}_{12}}\,\, {\rm Tr}_{N}
\left[{\cal F}'\right]\, \in\, {\bf Z}
\label{1st}
\end{equation}
Specifically the resulting gauge field configuration has $k'=k$. As we are 
integrating over $R^{2}_{12}$ only and not over $T^{2}$, equation (\ref{1st}) 
holds pointwise on $T^{2}$. In particular it still holds if we set 
$x_{1}=x_{2}=0$. As discussed in the previous section, in this case we can 
simply identify the components of ${\cal A}'(\vec{0},y)$ 
along $R^{2}_{12}$ with the 
corresponding components of the original four-dimensional gauge connection 
$a$. This means that,      
\begin{equation}
\frac{1}{2\pi}\int_{R^{2}_{12}}\,\, {\rm Tr}_{N}
\left[f\right]\, = k
\label{1stb}
\end{equation}
where $f$ is the curvature of the four-dimensional gauge connection $A$. 
In the four-dimensional theory, the first Chern class (\ref{1stb}) 
also coincides with the total $U(N)$ magnetic flux in the $(1,2)$-plane. 
Further, as the original $U(m)$ 
instanton solution is localised in the transverse 
$T^{2}_{\Theta}\times R^{2}$, each of the Fourier coefficients of its 
field strength and energy density is localized in the commutative coordinates 
$y_{1}$ and $y_{2}$, near the center of the 
instanton in $R_{12}^{2}$. 
Thus the resulting configuration in the four dimensional theory 
corresponds to a magnetic string carrying $k$ units of magnetic 
flux. In particular a single instanton corresponds to a string carrying 
the minimal quantum of magnetic flux. 
The core size of the flux tube is dictated by the scale-size $\rho$ 
of the corresponding instanton.
\paragraph{}
A more detailed picture of the resulting flux tubes can be gained by using a 
few simple facts about Yang-Mills instantons (see for 
example \cite{PREP}). To make the discussion as 
simple as possible, we work on  
square torus by setting $|\alpha_{1}|=|\alpha_{2}|=|\alpha|$. In this case  
we can identify four relevant length-scales in ascending order,  
\paragraph{}
$L_{1}=\sqrt{G_{6}^{2}m}=\sqrt{g^{2}N}/|\alpha|$. 
The length scale set by 't Hooft coupling of the six dimensional gauge theory.
\paragraph{}
$L_{2}=1/|\alpha|$. The lattice spacing.  
\paragraph{}
$L_{3}=\sqrt{\vartheta}=\sqrt{2\pi n}/|\alpha|$. 
The length-scale at which non-commutativity becomes significant. 
\paragraph{}
$L_{4}=n/2\pi|\alpha|$. The compactification radius. 
\paragraph{}
First we note that in the large-$n$ semiclassical regime 
we are considering ($n>>1$, $g^{2}N<<1$) 
all of these scales are well-separated: $L_{1}<<L_{2}<<L_{3}<<L_{4}$. 
As $L_{2}>>L_{1}$, the six-dimensional theory is weakly coupled on scales 
larger than the lattice spacing justifying a 
classical analysis. As noted above, the six-dimensional 
continuum effective action is valid for describing instantons with 
scale-size $\rho>>L_{2}$. 
In a non-commutative theory instantons typically 
have a minimum scale size of order the non-commutativity length scale 
$\rho_{min}\sim L_{3}>>L_{2}$. Thus all the low-energy action is valid for all 
the solutions we will consider. As above, a long 
string corresponds to a Yang-Mills instanton on the non-commutative space 
$T^{2}_{\Theta}\times R^{2}_{12}$. Such objects have been studied before 
and a version of the Nahm transform exists for constructing the corresponding 
moduli space of solutions \cite{Sethi}. 
However, for the present purposes a 
much simpler description will suffice. For $n>>1$, there is 
large heirarchy between the length scale set by the non-commutativity and the 
radius of compactification. In fact we have 
$L_{4}/L_{3}\sim \sqrt{n}\rightarrow \infty$. 
We will assume\footnote{Note that the results of \cite{Sethi} are 
consistent with this assumption. In the case where 
Wilson lines are turned off, the resulting 
moduli space of a single instanton,  
with fixed COM on non-commutative $R^{2}\times T^{2}$ 
is the same as in the non-compact case.} that instantons 
scale-size $\rho<<L_{4}$ exist 
and are well approximated by instantons on the non-compact Euclidean 
space $R^{2}_{\vartheta}\times R^{2}_{12}$ about which much more is known. 
\paragraph{}        
We begin with the case $m=1$ where the $U(N)$ gauge symmetry is broken 
down to $U(1)$. In this case we find a non-commutative $U(1)$ instanton 
with a fixed scale-size 
$\rho\sim L_{3}=\sqrt{2\pi n}/|\alpha|$. As noted above, 
this length-scale is larger than the lattice spacing by a factor of $\sqrt{n}$ 
justifying the use of continuum effective field theory. The presence of 
magnetic flux tubes of fixed size indicates that magnetic charges are 
confined in the large-$n$ limit. In this limit the central 
$U(1)$ factor in the gauge group effectively decouples and we can think of 
these strings as magnetic vortices carrying an N-valued topological 
charge corresponding 
to $\pi_{1}[SU(N)/{\bf Z}_{N}]={\bf Z}_{N}$. 
\paragraph{}
Extending the discussion to the Higgs branch ${\cal H}_{m}$ where the 
unbroken gauge group is $U(m)$, we must consider $U(m)$ instantons on 
$R^{2}_{\vartheta}\times R^{2}_{12}$. The occurence of a variable 
scale-size is related to the fact that the gauge group is only partially 
broken. Magnetic flux in the unbroken $U(m)$ is not confined and should 
therefore be able to spread out. Increasing the scale size of the 
instanton corresponds to increasing the core-size of the corresponding 
magnetic flux tube. We interpret this as the spreading out of the 
flux in the unbroken part of the gauge group. This interpretation is 
reinforced by recalling that the $U(m)$ instanton is not invariant under 
global $U(m)$ gauge rotations. Instead it has a fixed orientation inside the 
gauge group and the corresponding gauge orbit appears as the coset 
$U(m)/U(m-2)$ in its moduli space. 
\paragraph{}
Finally it is interesting to consider what happens when we move onto the 
Coulomb branch of the low-energy $U(m)$ theory. In this case the gauge group 
is further Higgsed down to $U(1)^{m}$. The effect of introducing such a VEV 
has on Yang-Mills instanton is well-known. It introduces a potential 
on the instanton moduli space. 
In the case of a single instanton we find a potential 
$V\sim v^{2}\rho^{2}$, where $v$ is the scale of the VEV, 
which tends to force the instanton to zero-size. In the non-commutative 
case we instead find a potential which looks like 
$v^{2}\rho^{2}$ for large $\rho$ but also has a nontrivial minimum at 
fixed size $\rho\sim L_{3}$. Thus we again end up with flux-tubes 
of fixed core-size and the confinement of the magnetic charge.
\paragraph{}
In the weak-coupling regime discussed above, the mass-scale $\sqrt{T}$ 
set by the string tension is large compared to the masses of the elementary 
quanta of the theory and the strings appear 
as classical solitons in the low-energy effective action. A 
semiclassical quantisation of these objects would involve 
studying a two-dimensional non-linear $\sigma$-model with ${\cal N}=(4,4)$ 
supersymmetry with the instanton moduli space as the target. On a branch 
with unbroken gauge-group $U(m)$ for $m>1$, the core-size of the string is a 
modulus of the solution and a semiclassical quantization of the string 
yields a continuous spectrum of states in six dimensions. On the branches 
discussed above where the gauge symmetry is further broken to $U(1)^{m}$, 
this modulus is lifted and the relevant $\sigma$-model now includes a 
potential term given by the norm of a Killing vector on the moduli space 
\cite{LT}. The potential has isolated minima and we would therefore expect 
normalizable string winding modes and a discrete spectrum of excited states.  
\paragraph{}
The most interesting questions concerning the strings discussed above 
relate to their behaviour away from weak coupling. As we have argued above, 
the semiclassical formula for the string tension 
$T=8\pi^{2}|\alpha_{1}||\alpha_{2}|/g^{2}n$ becomes exact in the 
large-$N$ limit. This suggests that there should be interesting regimes 
in the parameter space where the strings become light and the 
effective description should be a string theory, 
rather than a local quantum field theory. In general such a description is 
only useful if the strings are weakly interacting. In Section 9, 
we will suggest that, at least in one regime of parameters, this is indeed 
the case.     
\section{S-duality and Electric Confinement}
\paragraph{}
The $SL(2,{\bf Z})$ duality transformations (\ref{sl2z}) 
described in Section 2, 
relate theories with 
different values of the parameters $\tau$, $\beta$ and 
$\kappa=\kappa_{cr}[\tau,\beta]$. In particular they map vacuum states of a 
theory with one set of values of these parameters onto vacuum states of a 
theory with another set of values. Vacua on 
Coulomb or Higgs branches are characterized by non-zero VEVs for the chiral 
operators $u_{(k_{1},k_{2},k_{3})}$ discussed in Section 3. In the 
${\cal N}=4$ theory these operators are parts of chiral primary multiplets 
transforming with fixed modular weights $(k/2,k/2)$ where 
$k=k_{1}+k_{2}+k_{3}$ \cite{Int,ADK}. 
As discussed in Appendix A, the results of \cite{DHK} 
indicate that these operators retain 
the same weights under the modified $SL(2,{\bf Z})$ 
transformation (\ref{sl2z}) of the $\beta$-deformed theory.  
Given the VEVs of these operators in a vacuum of one 
theory, the transformation rule (\ref{transform}) specifies 
their VEVs in a corresponding vacuum state of the dual theory 
\paragraph{}
We now want to apply the S-duality transformation (\ref{sl2z}) 
given above to the 
$U(N)$ theory with $\beta=2\pi/n$ (where $N=mn$) on its Higgs branch 
${\cal H}_{m}$. Specifically we will consider the theory in the vacuum state 
(\ref{vacm}) parametrised by three complex numbers 
$\alpha_{1}$, $\alpha_{2}$ and $\alpha_{3}$. 
According to Eq (\ref{sl2z}), the dual theory will have deformation parameter 
$\tilde{\beta}=2\pi/n(c\tau_{R}+d)$. The resulting vacuum state can be 
described in terms of the gauge invariant chiral operators 
$u_{(k_{1},k_{2},k_{3})}$ defined in Section 3. 
As in the vacuum (\ref{vacm}), 
these operators will vanish in the dual vacuum unless {\em either}
\begin{eqnarray}
(k_{1},k_{2},k_{3})=(0,0,0) \qquad{} {\rm mod}\,\,\,n & \qquad{} {\rm or} 
\qquad{} & k_{1}=k_{2}=k_{3} \qquad{} {\rm mod}\,\,\,n \nonumber \\
\label{chiralvevs1b}
\end{eqnarray}     
In the non-vanishing case we have
\begin{eqnarray}
\langle u_{(k_{1},k_{2},k_{3})} \rangle & \qquad{}  = \qquad{}  & 
\exp (i\tilde{\nu}_{(k_{1},k_{2},k_{3})})
\tilde{\alpha}_{1}^{k_{1}}\tilde{\alpha}_{2}^{k_{2}}
\tilde{\alpha}_{3}^{k_{3}} \nonumber \\
\label{chiralvevs2b}
\end{eqnarray}
with $\tilde{\alpha}_{i}=\alpha_{i}|c\tau_{R}+d|$ for $i=1,2,3$ 
where $\exp(i\tilde{\nu}_{(k_{1},k_{2},k_{3})})$ is an unimportant 
phase.  
\paragraph{}
To simplify our discussion we will focus for a moment on the 
`S' generator of $SL(2,{\bf Z})$ and work with $n>>1$ so that 
$\beta=2\pi/n<<1$ and we can work to first order in the perturbation. 
As explained in Section 2, we then have $\kappa_{cr}=1+O(1/n^{2})$ and 
which gives $\tau_{R}=\tau+O(1/n^{2})$. We will also set the vacuum 
angle $\theta=0$. In this case the parameters of the theory transform as,  
\begin{equation}
g^{2} \rightarrow \tilde{g}^{2}_{R}=
\frac{16\pi^{2}}{g^{2}}  
\qquad{} \qquad{} \beta=\frac{2\pi}{n}\rightarrow \tilde{\beta}
=\frac{ig^{2}}{2n}
\label{sl2z4}\
\end{equation}
and the dual vacuum has VEVs specified by 
(\ref{chiralvevs1b},\ref{chiralvevs2b}) with 
$\tilde{\alpha}_{i}=\alpha_{i}(4\pi/g^{2})$ for $i=1,2,3$. 
Several features of this result are worthy of comment; 
\paragraph{}
{\bf 1:} The duality transformation has mapped a theory where the parameter 
$\beta/2\pi$ takes the rational real value $1/n$ has been mapped to  
one where this parameter takes an imaginary value $ig^{2}/4\pi n$.  
As we start from a three-parameter family of vacua on 
${\cal H}_{m}$, the image is a three-parameter family of vacua in the 
dual theory. In generic vacua, chiral operators such as 
$u_{(n,0,0)}$, $u_{(0,n,0)}$ and $u_{(0,0,n)}$ all have non-zero VEVs.  
Our classical analysis showed that 
branches like these, on which more than one of the three scalar fields 
$\Phi_{i}$ have a VEV, are only possible for real, rational values of 
$\beta/2\pi$. Thus the vacua in question lie on a new branch 
$\tilde{\cal H}_{m}$ of the dual quantum theory which is not 
visible classically.  
\paragraph{}
{\bf 2:} As the branch ${\cal H}_{m}$ exists for all values of the gauge  
coupling in the theory with $\beta=2\pi/n$, we deduce that the 
new branch $\tilde{\cal H}_{m}$ exists for all values of the 
renormalized 
gauge coupling $\tilde{g}^{2}_{R}$ in the $\beta$-deformed theory with 
deformation parameter $\tilde{\beta}=
8\pi^{2}i/\tilde{g}^{2}_{R}n$     
\paragraph{}
{\bf 3:} When the original Higgs phase theory is weakly coupled, 
$g^{2}<<1/N$, the dual 
gauge theory has ultra strong coupling $\tilde{g}_{R}^{2}>>N$. Note however 
that the converse is not true. When the Higgs theory has very strong  
coupling $g^{2}>>N$, the dual theory has a very small gauge coupling 
$\tilde{g}_{R}^{2}<<1/N$. However the coefficients appearing in front of the 
cubic scalar interactions in the superpotential are 
$\exp(\pm i\beta/2)=\exp(\mp 4\pi^{2}/\tilde{g}_{R}^{2}n)$. These lead to 
exponentially large quartic self-interactions of the scalars in this limit. 
Thus the dual theory is {\em never} weakly coupled and a semiclassical 
analysis of its vacuum structure is not reliable. Because of this, 
the existence of 
the new branch $\tilde{\cal H}_{m}$ does not lead to a contradiction.     
\paragraph{}
{\bf 4:} The $\beta$-deformed 
theory can be thought thought of as a small perturbation of the 
${\cal N}=4$ theory provided that 
$|\beta|=2\pi/n<<1$. 
In the S-dual vacuum the relevant condition becomes
$|\tilde{\beta}|=g^{2}/2n<<1$. Rewriting this in terms of the dual coupling 
we have $\tilde{g}^{2}_{R}n\simeq \tilde{g}^{2}n>>1$. Hence, when 
both $n>>1$ and $\tilde{g}^{2}n>>1$ hold, both theories are close to the 
${\cal N}=4$ point. In this regime, the S-duality between the two theories 
can be established using only the linearized analysis of Section 2.       
\paragraph{}
{\bf 5:} The action of the 
full S-duality group produces new branches in an infinite number of 
theories with deformation parameter $\tilde{\beta}$ and 
renormalized gauge couplings $\tilde{\tau}_{R}$ related 
as $\tilde\beta=2\pi(c\tilde{\tau}_{R}-a)/n$. None of the new branches is 
visible semiclassically.   
\paragraph{}
We can provide further evidence for the existence of the new branches 
directly from the analysis of chiral condensates in the massive 
version of the $\beta$-deformed theory with gauge group $SU(N)$. 
This theory has 
classical superpotential ${\cal W}+\Delta{\cal W}$ where the new term 
$\Delta{\cal W}$ is given in (\ref{rel}). The theory has an $SL(2,{\bf Z})$ 
multiplet of massive vacua in various Higgs and confining phases. The vacua 
are labelled by three integers $r$, $s$ and $t$ with $N=rs$ and 
$t=0,2,\ldots r-1$. Each vacuum is 
characterised by a different effective coupling 
$\hat{\tau}=(r\tau_{R}+t)/s$. The exact superpotential in each of these vacua 
was calculated in \cite{DHK}. For the $SU(N)$ theory the result is, 
\begin{equation}
{\cal W}_{\rm eff}=\frac{rN\mu M^{2}}{2\kappa^{2}\sin\beta}\, 
\frac{\theta^{'}_{1}\left(\frac{r\beta}{2}|\hat{\tau}\right)}
{\theta_{1}\left(\frac{r\beta}{2}|\hat{\tau}\right)}
\label{resultdhk}
\end{equation}   
\paragraph{}
This result allows us to calculate the vacuum expectation values 
of chiral operators in these vacua. Two operators we will focus on are, 
\begin{eqnarray} 
u=u_{(0,0,2)}=\frac{1}{N}{\rm Tr}_{N}\left[ \Phi^{2}_{3}\right] & 
\qquad{} \,\,\, & w=u_{(1,1,1)}=
\frac{1}{N}{\rm Tr}_{N}\left[\Phi_{1}\Phi_{2}\Phi_{3}\right] \nonumber \\
\label{uw}
\end{eqnarray}
By standard Ward identities, we can calculate the VEVs of these operators 
by taking appropriate derivatives of the superpotential, 
\begin{eqnarray} 
\langle u \rangle =\frac{1}{N} 
\frac{\partial{\cal W}_{\rm eff}}{\partial\mu} & 
\qquad{} \,\,\, & \langle w \rangle 
=-\frac{\exp\left(-\frac{i\beta}{2}\right)}{N}\, 
\left[\frac{\partial{\cal W}_{\rm eff}}{\partial\beta}  
+\frac{i}{2}\frac{\partial{\cal W}_{\rm eff}}{\partial\kappa}                  \right]\nonumber \\
\label{uwvevs}
\end{eqnarray}
\paragraph{}
In the massless the limit 
$\mu\rightarrow 0$, $M\rightarrow 0$, vacua of the massive theory 
go over to vacua of the massless 
$\beta$-deformed theory. Of course it is not necessarily true that all vacua 
of the massless theory can be obtained fom the very specific deformation we 
consider here. Fortunately, at least for the $SU(N)$ theory, we will be 
able to construct a one-parameter family of vacua on each of the 
the new branches described above\footnote{Unfortunately, there are no 
vacua of the massive $U(N)$ theory with similar limiting behaviour. 
Hence we cannot use the same arguments to demonstrate the existence of 
the new confining branches in this case. However the arguments given above, 
based on the S-duality of the massless $\beta$-deformed theory, apply equally 
to the $U(N)$ and $SU(N)$ theories.}.           
\paragraph{}
We begin by translating the vacuum structure of the massless theory into 
predictions for the condensates of the operators $u$ and $w$ defined above.    
For generic values of $\beta$, 
the massless theory only has a Coulomb branch. 
In some of these vacua $\langle u \rangle \neq 0$, but 
$\langle w \rangle=0$ in all vacua. On the other hand, when $\beta=2\pi/n$ 
with $N=nm$, the massless theory develops a Higgs branch ${\cal H}_{m}$. 
On this branch generic vacua have $\langle w \rangle \neq 0$, but 
$\langle u \rangle$ vanishes in all cases. In this Section, we have argued 
using S-duality, that the theory has an $SL(2,{\bf Z})$ multiplet of 
new branches all occuring at different values of $\beta$. As the operators $u$ and $w$ transform with definite modular weights, the generic vacua on 
{\em all} of these branches have $\langle w \rangle \neq 0$ and 
$\langle u \rangle=0$. In particular $\langle w \rangle$ can take an 
arbitrary complex value on each of these branches 
These branches also have other non-zero 
condensates, but to calculate these in the massive theory, 
we would need to extend the analysis of \cite{DHK} by 
adding sources for the corresponding operators to the 
superpotential. 
\paragraph{}
We will now attempt to recover this vacuum structure of by taking the 
massless limit $\mu\rightarrow 0$, $M\rightarrow 0$.
From (\ref{resultdhk}) and (\ref{uwvevs}), the VEVs of $u$ and $w$ 
are proportional to 
$M^{2}$ and $\mu M^{2}$ respectively and so naively they vanish in this limit. 
This conclusion is unavoidable, except for special points where the 
effective superpotential (\ref{resultdhk}) diverges. The ratio of 
$\theta$-functions appearing in (\ref{resultdhk}) has a simple pole when the 
argument $r\beta/2$ vanishes. It is also quasi-periodic in this 
parameter with periods $\pi$ and $\pi\hat{\tau}$, thus has one pole in each 
period parallelogram. In vacua with $t=0$, we have 
$\hat{\tau}=r\tau_{R}/s$ where 
$N=rs$, and the simple poles are located at, 
\begin{equation}
\beta=\frac{2\pi p}{r}+ \frac{2\pi q\tau_{R}}{s}=\beta_{cr}(p,q|r,s)      
\label{poles}
\end{equation}
for arbitrary integers $p$ and $q$. Remarkably, these 
singular points coincide precisely with all the values of $\beta$ 
where the arguments given above predict new branches. In particular, 
the Higgs branch ${\cal H}_{m}$ occurs at, 
$\beta=2\pi/n=\beta_{cr}(1,0|n,m)$. Restoring the integer $t=0,1\ldots r-1$ 
we find a pole corresponding to a Higgs branch for $\beta=2\pi l/N$ where 
$l$ can be any integer.  
The dual magnetic branch $\tilde{\cal H}_{m}$ 
occurs at $\beta=2\pi\tau_{R}/n=\beta_{cr}(0,1|m,n)$. 
Near each critical point 
we have, 
\begin{equation}
{\cal W}_{\rm eff}  \sim  \mu M^{2}\, \frac{1}{(\beta-\beta_{cr})}       
\label{wscale}
\end{equation}
Using (\ref{uwvevs}) we find, 
\begin{eqnarray}
\langle u \rangle \, \sim \,  M^{2}\, \frac{1}{(\beta-\beta_{cr})} & \qquad{} 
&  \langle w \rangle \, \sim \,  \mu M^{2}\, \frac{1}{(\beta-\beta_{cr})^{2}} 
\nonumber \\
\label{uwscale}
\end{eqnarray}
\paragraph{}
We now want to investigate the massless $\beta$-deformed theories at these 
special values of $\beta$.  We begin by taking the 
limit $M\rightarrow 0$, $\beta\rightarrow \beta_{cr}$ holding $\mu$ fixed. 
Specifically we take the limit 
$\beta-\beta_{cr}=\epsilon\rightarrow 0$ with $M\sim \epsilon$. In this case 
we end up with vacua in which $\langle u \rangle=0$ and $\langle w \rangle$ 
can take an arbitrary complex value. One can also check 
that the other condensates calculated in \cite{DHK} remain finite in 
this limit. Finally we can take the limit 
$\mu\rightarrow 0$ holding $\langle w \rangle$ fixed. 
Thus we have confirmed the existence of 
a branch of vacua with $\langle w \rangle\neq 0$ for each critical value 
of $\beta$. Of course, our choice of limits now implies that these branches 
also exist in the theory with one non-zero mass (ie $M=0$ and $\mu\neq 0$).   
This is not obvious but we check it explicitly in Appendix B. 
\paragraph{}
In order to understand the physics of the S-dual branch 
$\tilde{\cal H}_{m}$, it is useful to recall the discussion of the 
Higgs branch ${\cal H}_{m}$ given in Section 3. 
We start from the undeformed ${\cal N}=4$ theory on the one-dimensional 
submanifold of the Coulomb branch denoted ${\cal C}_{1}$ (see discussion 
above Eq (\ref{mass1})). 
As above, we set the vacuum angle $\theta$ to zero for 
simplicity.  The spectrum of the theory at this point includes $N^{2}$ 
W-bosons with masses $M^{(ab)}$ as given in (\ref{mass1}) and 
magnetic monopoles with masses $(4\pi/g^{2})M^{(ab)}$, 
as well as the usual spectrum of dyons. At weak 
coupling $g^{2}<<1/N$, introducing a real non-zero 
value for $\beta$ leads to mass splittings in the $W$-boson multiplet 
with half of each multiplet getting $\beta$-dependent masses as 
given in (\ref{mass2}). This results in the appearence of massless 
electrically charged states when $\beta=2\pi/n$ where $N=mn$. At this point 
the theory develops a Higgs branch ${\cal H}_{m}$ on which these 
massless electric states condense breaking the $U(N)$ gauge group to $U(m)$.
As discussed in Section 3, neither the Higgs branch nor the Coulomb branch 
can be lifted by quantum effects. Similarly, the orbifold singularity 
where the two branches intersect cannot be resolved by quantum corrections 
and the two branches will continue to intersect for all values of the 
gauge coupling. Thus, although the masses of the $W$-bosons can be 
renormalized for $\beta\neq 0$, we expect that the singular point 
where they become massless will still exist in the full quantum theory   
\paragraph{}
At the classical level, the masses of the magnetic monopoles are unaffected 
by a non-zero value of $\beta$. However, in the presence of non-zero $\beta$, 
these states, like the W-bosons, are no longer BPS and we would 
certainly expect their masses to 
recieve $\beta$-dependent quantum corrections.
Under an S-duality transformation (\ref{sl2z4})  
the W-bosons and 
magnetic monopoles of the Coulomb branch theory are exchanged. 
The S-transformation also maps the theory discussed above with 
$\beta=2\pi/n$ to an S-dual theory with 
$\beta=8\pi^{2}i/\tilde{g}_{R}^{2}n$. Hence we can 
map the above statements about the weak-coupling spectrum of W-bosons 
in the first theory to a corresponding statement about the 
strong-coupling spectrum of magnetic monopoles in the second. 
In this way we can infer the existence of a point on the Coulomb branch  
of the dual theory with 
massless magnetically-charged states. By S-duality we also find a 
new branch of vacua $\tilde{\cal H}_{m}$ on which magnetically 
charged states condense 
breaking the dual magnetic gauge group down to a $U(m)$ subgroup. 
Similar arguments indicate the existence of points where various 
dyonic states becomes massless and condense leading to an $SL(2,{\bf Z})$ 
multiplet of new branches in oblique confining phases. 
\paragraph{}
Of course, all of this is very reminiscent of Seiberg-Witten theory 
\cite{SW1}. Even though we have only ${\cal N}=1$ 
supersymmetry in our 
case, the Coulomb branch action is characterized by an $N\times N$ 
matrix $\tau_{ij}$ of abelian couplings as in the ${\cal N}=2$ case. 
As usual low-energy electric-magnetic duality can be encoded by 
identifying this matrix with the period matrix of a genus $N$ Riemann 
surface. The relevant curve should be closely related to the one 
given in \cite{Tim}.    
\paragraph{}
For the $SU(N)$ theory, another heuristic way of understanding 
the phase in which each vacuum is 
realised is via the massive deformation discussed above. In particular, we 
obtained a point on each new branch as a limit of a massive vacuum. The 
massive vacua in question reduce to the known vacua of the 
${\cal N}=1^{*}$ theory for $\beta=0$. The classical Higgs branch 
${\cal H}_{m}$ is continuously related to the ${\cal N}=1^{*}$ vacuum 
with effective coupling $\hat{\tau}=n\tau_{R}/m$. This vacuum is visable  
classically and has unbroken gauge symmetry $SU(m)$. This matches the 
low-energy gauge symmetry on ${\cal H}_{m}$. In the ${\cal N}=1^{*}$ 
theory the unbroken $SU(m)$ becomes strongly coupled and is 
confined with a mass gap in the IR. On the Higgs branch ${\cal H}_{m}$, 
the unbroken gauge symmetry is realised in a non-abelian Coulomb phase.   
Thus, in the simultaneous massless 
and $\beta\rightarrow\beta_{cr}=2\pi/n$ 
limit discussed above, the mass gap goes to zero 
and the unbroken $SU(m)$ is deconfined. 
\paragraph{}
The dual magnetic branch 
$\tilde{\cal H}_{m}$ corresponds to an ${\cal N}=1^{*}$ vacuum 
with $\hat{\tau}=m\tau_{R}/n$. This is normally interpreted as a vacuum 
in which the gauge group is Higgsed from $SU(N)$ down to $SU(n)$ which is 
then confined in the IR \cite{VW}. In 't Hoofts classification of 
massive phases, this is indistinguishable from a phase in which $SU(N)$ is 
confined down to $SU(m)$ which is then Higgsed in the IR. In fact one can 
move smoothly between these phases by varying the 
relative mass scales of the electric and magnetic condensates 
in the vacuum.  As we take the massless limit, the low-energy 
$SU(m)$ symmetry is restored leaving a theory on $\tilde{\cal H}_{m}$ 
which has $SU(N)$ confined down to $SU(m)$ as expected.     
\paragraph{}
On the Higgs branch ${\cal H}_{m}$ we found that chromomagnetic charges are 
confined by magnetic flux tubes of finite tension which become 
BPS-saturated in the large-$n$ limit. The S-dual statement is that on the 
magnetic Higgs branch $\tilde{\cal H}_{m}$, we should find the 
confinement of chromoelectric charge by  
electric flux tubes of finite tension, 
\begin{equation}
\tilde{T}=\frac{8\pi^{2}|\tilde{\alpha}_{1}|
|\tilde{\alpha}_{2}|}{\tilde{g}_{R}^{2}n}
\label{t2}
\end{equation}
Thus we interpret the dual Higgs branch $\tilde{\cal H}_{m}$ as one in which 
the original electric gauge group is confined down to a $U(m)$ subgroup which 
remains in a non-abelian Coulomb phase. Moving onto the 
Coulomb branch of the effective theory yields interesting vacua in which the 
gauge group is further confined down to $U(1)^{m}$. 
\paragraph{}
The regime of 
ultra-strong coupling $\tilde{g}^{2}_{R}>>N$ on this confining branch 
is dual to the weakly-coupled theory on the Higgs branch studied above. 
In this regime, long chromoelectric strings emerge as BPS saturated solitons 
in a low-energy effective theory with sixteen supercharges. As in the 
Higgs phase, an obvious question is whether the theory has a regime where 
the strings become light. In the Higgs phase, the BPS formula 
(\ref{tension}) becomes exact in the $N\rightarrow \infty$ limit 
(with the gauge coupling $g^{2}$ held fixed). Thus, by S-duality, 
(\ref{t2}) gives the exact tension of the confining strings in the 
large-$N$ limit. This again suggests the existence of interesting 
regimes of parameter space where the confining strings become light. 
One such regime is a 't Hooft large-$N$ limit in which 
$N\rightarrow \infty$ with $\tilde{g}_{R}^{2}n$ held fixed and large.
This is very suggestive because 
of the standard lore about confining gauge theories in the 
't Hooft limit: one expects an infinite tower of glueball states 
corresponding to the fluctuations of a light string. 
In Section 9, we will make a concrete proposal for the dual 
string theory for the confining branch of the $\beta$-deformed theory.       
\section{String Theory Realisations}
\paragraph{}
As in many other examples, string theory can provide important insights about 
the dynamics of the gauge theory described above. The first step is to realise 
the theory in question on the world-volume of a suitable D-brane 
configuration. The undeformed ${\cal N}=4$ theory with gauge group $U(N)$ 
can be realised on the world volume of $N$ parallel D3 branes in 
Type IIB string theory. The four dimensional gauge coupling is given in terms 
of the string coupling, $g_{s}$, as $g^{2}=4\pi g_{s}$ and the 
$\theta$-angle is related to the RR scalar field $C_{(0)}$ of the 
IIB theory so that we have, 
\begin{equation}
\tau=\frac{4\pi i}{g^{2}}+ \frac{\theta}{2\pi}= \frac{i}{g_{s}}+ 
C_{(0)} 
\label{tau}
\end{equation}
With this identification the $SL(2,{\bf Z})$ duality of 
the ${\cal N}=4$ theory is 
mapped onto the S-duality of the IIB theory.  To decouple the 
gauge theory, both from gravity and from the excited states of the 
open string we must take the limit $\alpha'\rightarrow 0$ holding 
$g_{s}$ fixed. We must also rescale the spatial coordinates transverse 
to the brane keeping the masses of open strings stretched between the 
branes fixed in the $\alpha'\rightarrow 0$ limit. When the 't Hooft coupling 
$\lambda=g^{2}N$ is small, the gauge theory is weakly coupled. 
If it is large, the gauge theory is strongly coupled and has a 
dual description in terms of 
IIB string theory on the near horizon geometry of the D3-branes, 
which is $AdS_{5}\times S^{5}$ with $N$ units of Ramond-Ramond 
five-form flux through $S^{5}$ \cite{Mal}.      
\paragraph{}
Introducing the marginal $\beta$-deformation corresponds to turning on a 
particular background flux for the complexified three-form 
field strength of the theory, 
\begin{equation}
G_{(3)}=F_{(3)}-\tau H_{(3)}
\end{equation}
The particular mode which must be turned on can be deduced from the standard 
dictionary between field theory operators and SUGRA fields provided by 
the AdS/CFT correspondence. Specifically the SUGRA fields are expanded in 
appropriate spherical harmonics of the $SO(6)\simeq SU(4)$ rotational 
symmetry of the directions transverse to the branes and these modes are 
identified with chiral operators transforming in the corresponding 
representations of the $SU(4)$ R-symmetry of the ${\cal N}=4$ theory. 
As discussed in Section 2, the chiral operator coresponding to the 
$\beta$-deformation is part of 
the ${\bf 45}$ of $SU(4)_{R}$. The string theory realisation of the 
$\beta$-deformation therefore involves turning on the corresponding spherical 
harmonic of $G_{(3)}$ with (complex) field strength 
proportional to $\beta$.    
\paragraph{}
As we saw in Section 3, turning on an arbitrarily 
small value of $\beta$ can have a significant effect. Specifically when $N=nm$ 
with $n>>1$, choosing the value $\beta=2\pi/n$ leads to the 
existence of a new Higgs branch ${\cal H}_{m}$ on which the $U(N)$ 
gauge group is spontaneously broken to $H=U(m)$. As usual we can 
deduce the corresponding distribution of D3 branes by looking at the 
eigenvalues of the three complex scalar fields $\Phi_{i}$. We consider the 
vacuum with scalar VEVs given by (\ref{vacm}). The three matrices 
$U_{(n)}$ and $V_{(n)}$ and $W_{(n)}$ appearing in (\ref{vacm}) 
have the same eigenvalues: the $n$ distinct 
$n$'th roots of unity. We define complex coordinates on the 
six dimensions transverse to the branes as, 
\begin{equation}
\begin{array}{ccc}
z_{1}= x_{1}+ix_{4}=
\rho_{1}e^{i\psi_{1}} & z_{2}=x_{2}+ix_{5}=\rho_{2}e^{i\psi_{2}} & 
z_{3}=x_{3}+ix_{6}=\rho_{3}e^{i\psi_{3}} 
\end{array}  
\label{complex}
\end{equation}
The simplest case is when only two of the three fields have VEVs. 
Hence we set $\alpha_{3}=0$ in (\ref{vacm}). For large-$n$, the D3-branes 
are uniformly distributed on a square torus $T^{2}(r_{1},r_{2})$ 
in $R^{6}$ defined by 
$\rho_{1}=r_{1}=|\alpha_{1}|(2\pi \alpha')$, 
$\rho_{2}=r_{2}=|\alpha_{2}|(2\pi \alpha')$, $z_{3}=0$. Reintroducing 
non-zero $\alpha_{3}$, has the effect of slanting the torus. In this  
more general case the resulting two-dimensional torus is defined by 
$\rho_{i}=r_{i}=|\alpha_{i}|(2\pi \alpha')$ for $i=1,2,3$ with the extra 
condition, 
\begin{equation}
\psi_{1}+\psi_{2}+\psi_{3}=\delta_{1}+\delta_{2}+\delta_{3}
\label{extra}
\end{equation}
where $\alpha_{i}=|\alpha_{i}|\exp(i\delta_{i})$. In the following we will 
focus on the simplest case of a square torus, setting $\alpha_{3}=0$, unless 
otherwise stated.   
\paragraph{}
The brane configuration described above carries 
$N$ units of D3-brane charge. The fact that 
$\Phi_{1}$ and $\Phi_{2}$ do not commute implies that it also carries 
non-zero moments of D5-brane charge. 
This follows from the usual identification of the commutator of the 
scalar fields on a Dp-brane world-volume with D(p$+2$)-brane density. 
In fact, the VEVs of the scalar fields coincide precisely 
with those occuring in the standard M(atrix) theory construction of $m$  
toroidally wrapped D(p$+2$)-branes \cite{WATI}. The 
corresponding M(atrix) Lagrangian is that of the undeformed 
${\cal N}=4$ theory. In this context, the configuration is 
unstable at finite $n$, only becoming a true ground state in the 
$n\rightarrow \infty$ limit. In the present case, the presence of the 
$\beta$-deformation of the ${\cal N}=4$ theory has the effect of rendering 
the configuration stable.
\paragraph{}
At large-$n$, the brane configuration corresponding to our chosen vacuum 
therefore consists of $m$ D5-branes wrapped on the torus $T^{2}(r_{1},r_{2})$. 
This is completely analogous to the 
appearance of spherically wrapped D5 branes in the string theory 
realisation of the ${\cal N}=1^{*}$ theory \cite{PS}. 
In particular, the ${\cal N}=1^{*}$ theory has a vacuum in which $U(N)$ is 
broken to $U(m)$, which corresponds to $m$ D5 branes wrapped on 
a fuzzy sphere. Indeed, for the $SU(N)$ theory, 
we argued above that this vacuum can actually be 
deformed\footnote{The appearance of extra massless modes as we approach the Higgs branch vacuum seems consistent with the fact that a sphere cannot be 
{\em continuously} deformed into a torus.} into a vacuum on the Higgs branch 
${\cal H}_{m}$ by turning on the deformation parameter and taking the 
masses to zero. However there are also some important differences between the 
two cases. As mentioned in the introduction, toroidal compactification 
preserves supersymmetry while spherical compactification 
breaks most of it. Another important difference is that the ${\cal N}=1^{*}$ 
vacua are isolated while in the massless $\beta$-deformed case 
we have a continuous degeneracy of vacua. Correpondingly, the fuzzy sphere's 
appearing in the ${\cal N}=1^{*}$ case have fixed radii set by the mass 
parameters, while the fuzzy torii discussed above have variable radii set by the Higgs branch VEVs.                    
\paragraph{}
The worldvolume theory on the toroidally-wrapped D5-branes 
is an ${\cal N}=(1,1)$ supersymmetric 
six-dimensional gauge theory with gauge group $U(m)$. Thus we can already 
see the emergence of the six-dimensional 
effective theory derived in the previous Sections.   
Compactification of this theory on a torus with periodic 
boundary conditions preserves all the supersymmetries.  
The presence of $N$ units of D3 brane 
charge is realised on the D5-brane world-volume a constant 
non-zero background for the Neveu-Schwarz two-form $B_{\rm NS}$. 
\begin{equation}
\int_{T^{2}}\, B_{\rm NS}= 2\pi n (2\pi \alpha')
\label{quantum}
\end{equation}
We introduce flat coordinates $\chi_{1}$ and $\chi_{2}$ on the torus 
with $0\leq \chi_{1}\leq 2\pi r_{1}$ and $0\leq \chi_{2}\leq 2\pi r_{2}$. 
In this basis the metric on the torus is the flat one 
$g_{ij}=\delta_{ij}$. The two-form field can be written as, 
 \begin{equation}
B_{\rm NS}=\frac{n}{2\pi r_{1}r_{2}}(2\pi\alpha') 
\, d\chi_{1}\wedge d\chi_{2} = 
\frac{n}{2\pi |\alpha_{1}||\alpha_{2}|}\, \frac{1}{(2\pi \alpha ')} 
\, d\chi_{1}\wedge d\chi_{2}                                                   
\label{bns}
\end{equation}
and the corresponding `flux' per unit area\footnote{
The word flux appears here in inverted commas because $B_{\rm NS}$ is a 
two-form potential rather than a field-strength.} of the torus is given by, 
\begin{equation}
{\cal B}= \frac{1}{4\pi^{2}r_{1}r_{2}}\int_{T^{2}}\, B_{\rm NS}=
\frac{n}{2\pi |\alpha_{1}||\alpha_{2}|}\, \frac{1}{(2\pi \alpha ')}
\label{flx}
\end{equation}
\paragraph{}
As usual, a non-zero value of $B_{\rm NS}$ on a D-brane 
world-volume induces non-commutativity. The low-energy theory on the 
D5 world-volume is the six-dimensional $U(m)$ theory compactified 
down to four dimensions on a non-commutative torus.         
To identify the parameters of this non-commutative gauge theory, 
we follow the procedure advocated by Seiberg and Witten 
\cite{SWNC}. The metric $g_{ij}$ 
introduced above is the metric on the torus as seen by closed strings. 
It is important to distinguish this metric from the effective metric 
$G_{ij}$ which appears in the open string S-matrix elements. 
Similarly we must distinguish between the closed string coupling $g_{s}$ and 
its open string counterpart $G_{s}$. The relation between the two sets of 
quantities is, 
\begin{eqnarray} 
G_{ij}= -\left(B_{\rm NS}g^{-1}B_{\rm NS}\right)_{ij} & \qquad \qquad{} 
& 
G_{s}=g_{s}{\rm det}^{-\frac{1}{2}}\left(B_{\rm NS}g^{-1}\right)
\nonumber \\
\label{open}
\end{eqnarray}
The first relation in (\ref{open}) implies that the radii of the torus 
as measured in the open string metric are, 
\begin{eqnarray}
R_{1}=B^{12}_{\rm NS}r_{2}= \frac{n}{2\pi |\alpha_{1}|} 
& \qquad{} \qquad{} & R_{2}= -B^{21}_{\rm NS}r_{1}= 
\frac{n}{2\pi |\alpha_{2}|}
\label{radii} 
\end{eqnarray}
The second relation in (\ref{open}) yields the six-dimensional  
coupling; 
\begin{equation}
G^{2}_{6}= G_{s}^{2}=\frac{g^{2}n}{|\alpha_{1}||\alpha_{2}|}
\label{g61}
\end{equation}
\paragraph{}
If we choose `open string' coordinates $\hat{x}_{1}$ and $\hat{x}_{2}$ on the 
torus with $0\leq \hat{x}_{1} \leq 2\pi R_{1}$ and 
$0\leq \hat{x}_{2} \leq 2\pi R_{2}$. These coordinates have 
non-vanishing commutator $[\hat{x}_{i},\hat{x}_{j}]=i\vartheta$ with 
$\vartheta=B_{\rm NS}^{-1}$. This gives $\vartheta_{ij}=\vartheta\varepsilon_{ij}$ 
with,  
\begin{equation}
\vartheta=\frac{n}{2\pi |\alpha_{1}||\alpha_{2}|}
\label{nc}
\end{equation}
Thus the parameters of the low-energy theory on the D5 branes are 
in precise agreement with the results of our earlier field theory 
analysis.              
\paragraph{}
This stringy derivation of the low-energy theory on the 
branes is valid when the closed string coupling 
is small: $g_{s}=g^{2}/4\pi<<1$. At large-$N$, this is a much 
weaker condition than the 
condition of small 't Hooft coupling, $g^{2}N<<1$ 
required to justify the semiclassical field theory analysis of Section 4. 
As in \cite{Mal}, 
it also includes a regime of strong 't Hooft coupling, where 
the worldvolume gauge theory has a dual description in terms of 
low-energy supergravity. The fact that the resulting 
parameters of the low-energy theory agree with the 
field theory results (\ref{table1},\ref{table2},\ref{table3}) is consistent 
with our claim that the Kaluza-Klein spectrum (\ref{mass3}) becomes 
exact at large-$N$. The supergravity description of the brane configurations 
described in this Section is studied in detail in \cite{D1}. 
\paragraph{}
In general, the gauge theory on D5 branes is still coupled both 
to gravity and to the excited states of the open string. In the present case 
a limit 
which decouples the worldvolume gauge theory from these extra degrees of 
freedom was identified in \cite{HOSJ}\footnote{See the $m=1$ case of the 
discussion of D5 branes given in Section 2.3 of this reference.}. 
The limit is one in which 
$\alpha'\rightarrow 0$ and ${\cal B}\rightarrow \infty$ with $g_{s}$ and 
$\alpha'{\cal B}$  held fixed. Here ${\cal B}$ is the 
background two-form `flux' per unit area defined in (\ref{flx}) 
above. The resulting theory reduces to a six-dimensional $U(m)$ gauge theory 
with non-commutativity in two space-like 
dimensions at low energies. At high-energies, it has a dual description 
in terms of IIB supergravity in the near-horizon geometry of the branes. 
For fixed $N$, the D5 brane decoupling 
limit is not the same as the decoupling limit which isolates the 
four-dimensional $\beta$-deformed theory on the original D3 branes.  
This does not present a puzzle because we know that the four-dimensional 
$\beta$-deformed theory should differ from its 
six-dimensional low-energy effective theory on 
length scales of order the lattice spacing. In Section 9, we will 
instead relate the decoupling limit for the D5 brane world-volume theory to 
a continuum limit in which the lattice spacing goes to zero.      
\paragraph{}
The brane configuration described above also 
yields a nice way of understanding the confinement of magnetic 
charge discussed in Section 5 \cite{PS}. 
We start from the ${\cal N}=4$ theory with gauge group 
$G=U(N)$ as realised on the world-volume of $N$ D3-branes. 
An external magnetic charge is represented as a semi-infinite D-string 
ending on the D3 brane. This corresponds to a single unit of non-abelian 
magnetic charge $k=1$ as measured by $k\in \pi_{1}[U(1)]={\bf Z}$. 
On the other hand, a single D-string parallel to the D3 world volume 
is absorbed onto the D3 worldvolume as one unit of magnetic flux through 
the transverse plane. As the ${\cal N}=4$ theory is in a non-abelian 
Coulomb phase the flux is not confined but spreads out as uniformly as 
possible
\paragraph{}   
After introducing the $\beta$-deformation with $\beta=2\pi/n$ and moving 
onto the Higgs branch ${\cal H}_{m}$, the $N$ D3 branes are uniformly 
spread on the torus $T^{2}(r_{1},r_{2})$ and we also have $m$ wrapped 
D5-branes. A D-string cannot end on a D5 brane 
but a D-string lying parallel to the D5 worldvolume can be 
absorbed as an instanton string in the six-dimensional world-volume theory 
on the D5s. The strings in question have 
a finite core-size set by the scale-size $\rho$ of the instanton. 
A D-string which approaches the D5 world-volume along a perpendicular 
cannot end but it may bend 
through $90^{0}$ and smoothly join onto a worldvolume string. 
If two D-strings of opposite orientation approach the D5 
worldvolume, the lowest energy configuration is one in which they are smoothly joined in this way by a worldvolume instanton string. This corresponds to 
the confinement of magnetic charges by a linear potential whose slope is 
set by the tension of the instanton string.   
\paragraph{}
For completeness we also consider the D-brane configurations which  
correspond to more general vacua in which the unbroken gauge group 
$U(m)$ is further broken to its Cartan subalgebra. Specifically 
we consider the vacua characterized as in (\ref{vac2}) by $m\times m$ 
diagonal matrices $\Lambda^{(i)}$ for $i=1,2,3$. We choose the eigenvalues 
as $\lambda^{(i)}_{s}=\alpha_{i}\mu^{(i)}_{s}$ for $s=1,2,\ldots,m$ where 
$\mu^{(i)}_{s}$ are real numbers. As above we set $\alpha_{3}=0$ for 
simplicity. Hence, at large $n$, the D3-branes are uniformly distributed on 
$m$ concentric torii $T^{2}(r^{(1)}_{s},r^{(2)}_{s})$ of radii 
$r^{(i)}_{s}=|\alpha_{i}|\mu^{(i)}_{s}(2\pi \alpha')$ for $i=1,2$. 
As above, we interpret these configurations as $m$ D5 branes wrapped on the 
corresponding concentric torii each carrying $n$ units of D3-brane charge. 
The light spectrum now includes W-bosons which correspond to open strings 
streched between torii with different radii. Extra massless states occur 
whenever pairs of torii coincide, corresponding to the restoration of 
a non-abelian subgroup of $U(m)$.   
\paragraph{}
So far we have restricted our attention to weak string coupling $g_{s}<<1$ 
but we would now 
like to consider what happens when the string coupling is increased. 
As discussed above, field theory considerations indicate that the 
Higgs branch cannot be lifted by quantum corrections. This suggests that 
the brane configuration described above remains stable as the string 
coupling is increased. As $g_{s}\rightarrow \infty$ we may use the S-duality 
of the IIB theory (which is equivalent to the field theory S-duality 
discussed above) to find a dual brane configuration in the weakly 
coupled string theory. Under S-duality the parameters of the IIB theory 
transform as, 
\begin{equation}
g_{s}\rightarrow \tilde{g}_{s}=\frac{1}{g_{s}} \qquad{} \qquad{} 
\alpha' \rightarrow \tilde{\alpha}'=g_{s}\alpha'
\label{2bsdual}
\end{equation}
The $m$ wrapped D5 branes are mapped onto $m$ wrapped NS5-branes. Using the 
above transformation laws we can rewrite the radii of the torus in terms of 
the S-dual variables as $r_{1}=|\tilde{\alpha}_{1}|(2\pi \tilde{\alpha}')$ 
and $r_{2}=|\tilde{\alpha}_{2}|(2\pi \tilde{\alpha}')$. The $N$ 
units of D3 brane charge are invariant under S-duality and can be realised on 
the NS5 worldvolume as a constant background for the Ramond sector 
two-form potential $B_{\rm RR}$.
More general Higgs vacua corresponding to D5 branes wrapped on concentric 
torii of different radii can also have S-duals with NS5 branes replacing the 
$D5$s. In this case, the spectrum includes massive `magnetic $W$-bosons' 
corresponding to D-strings stretched between the NS5-branes of 
different radii. 
In each of these vacua, we now have a natural explanation for the 
confinement of electric charges which is dual to the corresponding discussion 
of magnetic charges given above. External electric charges correspond to 
fundamental strings. These objects cannot end on an NS 5-brane, but 
instead may form a bound-state with it which corresponds to an 
instanton string in the world-volume gauge theory. Thus the instanton strings 
are identified as chromoelectric flux tubes in agreement with our 
earlier field theory discussion                                  
\paragraph{}
The theory also has complicated set of vacua in other phases. For example, we 
may act with the whole S-duality group on ${\cal H}_{m}$ to obtain 
branches of vacua in interesting 
oblique confining phases in which the unconfined 
low-energy gauge group is $U(m)$. These vacua correspond to $m$ 
$(p,q)$ 5-branes wrapped on $T^{2}(r_{1},r_{2})$. For each co-prime 
values of $p$ and $q$, the corresponding branch exists only for a particular 
value of $\beta$. Thus unlike the ${\cal N}=1^{*}$ case we never have any 
two of the branches coexisting in the same $\beta$-deformed theory. 
By moving onto the Coulomb branch of the low-energy theory we obtain a vacuum 
where the gauge group is Higgsed, confined or obliquely confined down to 
$U(1)^{m}$. These branches are realised as $m$ $(p,q)$ 5-branes wrapped on 
concentric torii.
\section{Towards a Continuum Limit}
\paragraph{}
In Section 4, we argued that the $\beta$-deformed theory on its Higgs branch 
${\cal H}_{m}$ is equivalent to a six-dimensional lattice gauge theory. 
The parameters of the six dimensional theory are the 
lattice spacings, $\varepsilon_{1}$ and $\varepsilon_{2}$, the 
radii of compactification, $R_{1}$ and $R_{2}$, as well as the six-dimensional 
coupling, $G_{6}$, and non-commutativity parameter $\vartheta$. These 
quantities are given in terms of the parameters of the 
four-dimensional theory in Equations 
(\ref{table1},\ref{table2},\ref{table3}) to which we now refer. 
\paragraph{} 
Given any lattice theory an obvious question is whether we can obtain an 
interacting continuum limit. In the present context, this means taking a 
limit in which $\varepsilon_{i}\rightarrow 0$, for $i=1,2$, with 
$G_{6}$ held fixed. In Sections 9.1 and 9.2 we will discuss two such 
limits. In the final subsection we discuss continuum limits of the theory in 
the S-dual Higgs/confining phases where the unbroken/unconfined gauge 
group is $U(1)^{m}$  
A more detailed discussion of the results presented in this 
Section is given in \cite{nd2}   
\subsection{A matrix theory limit}
\paragraph{}
To achieve a continuum limit with fixed non-zero 
six-dimensional coupling we can take $N\rightarrow \infty$ while holding 
$m=N/n$ and $g^{2}$ fixed. We therefore choose the scaling, 
\begin{eqnarray}
|\alpha_{1}| \sim \sqrt{n}\rightarrow \infty & \qquad \qquad & 
|\alpha_{2}| \sim \sqrt{n}\rightarrow \infty \nonumber \\
\label{scalings2}
\end{eqnarray}
This means that $G_{6}$ and $\vartheta$ remain fixed. With these 
choices the radii $R_{1}=n/2\pi|\alpha_{1}|$ and 
$R_{2}=n/2\pi|\alpha_{2}|$ grow like $\sqrt{n}$ and the torus 
decompactifies. These scalings are similar 
to the standard Matrix theory construction 
of $m$ D(p+2)-branes starting from $N$ Dp-branes \cite{Mat}.
Note however that the four dimensional 't Hooft coupling diverges in this 
limit: $g^{2}N\rightarrow \infty$. Thus we cannot justify the 
existence of the continuum limit purely on the basis of 
semiclassical field theory 
analysis. On the other hand, we have argued above that the whole spectrum 
of Kaluza-Klein states  
becomes BPS-saturated at large-$N$ and the the mass formula (\ref{mass3}) 
becomes exact. Thus the spectrum is consistent with the emergence of 
a continuum theory in six dimensions in the proposed continuum limit. 
\paragraph{}
We will now consider what kind of 
continuum theory could arise in such a limit. Naively, we would obtain a 
continuum six-dimensional non-commutative gauge theory.     
Of course six-dimensional gauge theories are non-renormalizable, so this 
cannot be the whole story: any continuum limit must involve some 
new physics in the UV. In the commutative case, 
there is only one known non-gravitational theory 
which reduces to six-dimensional ${\cal N}=(1,1)$ supersymmetric 
gauge theory at low energy. This is the decoupled theory on coincident 
NS5 branes of the Type IIB theory which is known as Little String Theory 
\cite{LST}. A more general theory\footnote{See the $m=1$ D5 brane case 
discussed in Section 2.3 of this reference as well as the 
S-dual discussion of NS 5 branes in Section 3.2} which reduces at low energies 
to a continuum six-dimensional $U(m)$ gauge theory with sixteen supercharges 
and non-commutativity in two spacelike dimensions was identified in 
\cite{HOSJ}. For $g^{2}<<1$, the UV behaviour of the 
theory is governed instead by 
IIB supergravity in the near-horizon geometry of the D5 branes with 
background $B_{\rm NS}$. The theory has one dimensionless parameter which 
controls the ratio of the non-commutative length 
scale and the six-dimensional gauge coupling. In a limit where this 
ratio goes to zero the theory reverts to the ordinary commutative  
Little String Theory.      
\paragraph{}
It is not hard to motivate the appearance of the continuum 
six-dimensional theory described above in the present context. 
In fact we already encountered the theory in question in Section 8, where 
it appeared as the Seiberg-Witten \cite{SWNC} decoupling limit of the 
world-volume gauge theory on the $m$ D5 branes in the string theory 
construction of the Higgs branch ${\cal H}_{m}$ at large $N$. The 
decoupling limit involved taking $\alpha'\rightarrow 0$ and 
${\cal B}\rightarrow \infty$ 
simultaneously while holding $g_{s}$ and $\alpha'^{2}{\cal B}$ 
fixed. As above, ${\cal B}$ is the 
background two-form `flux' per unit area of $B_{\rm NS}$ 
defined in (\ref{flx}) above.   
In terms of field theory variables the quantities held fixed are 
$g^{2}$ and $n/|\alpha_{1}||\alpha_{2}|$ respectively. 
At a fixed value of $N$, the decoupling limit for $\beta$-deformed 
${\cal N}=4$ theory realised on $N$ D3 branes is simply 
$\alpha'\rightarrow 0$. The fact that these limits are different reflects the 
fact that the $\beta$-deformed theory only agrees with the worldvolume 
theory on the D5 branes on lengthscales larger than the lattice spacing.  
However, if we combine the $\alpha'\rightarrow 0$ limit with the proposed 
continuum limit (\ref{scalings2}) in which the lattice spacing goes to zero, 
we have just the right scalings to 
decouple the six-dimensional gauge theory on the D5 branes. 
Thus we propose that, in the continuum limit described above, 
the $\beta$-deformed theory on its Higgs branch ${\cal H}_{m}$ 
is fully equivalent to the decoupled theory on the D5 branes. 
\paragraph{}
The resulting continuum theory in six dimensions 
has only one dimensionless parameter, the gauge coupling $g^{2}=4\pi g_{s}$. 
In the limit $g^{2}\rightarrow\infty$, the length-scale of non-commutativity 
goes to zero and the theory becomes ${\cal N}=(1,1)$ Little String Theory. 
In string theory this limit is best understood by performing an S-duality 
transformation (\ref{2bsdual}) 
which turns the $m$ D5 branes into $m$ NS5 branes 
with dual string coupling $\tilde{g}_{s}\rightarrow 0$ with 
$\tilde{\alpha}'$ held fixed. In field theory, 
the corresponding S-duality transformation yields a 
$\tilde{g}_{R}^{2}\simeq\tilde{g}^{2}\rightarrow 0$ 
limit of the theory on the magnetic 
Higgs branch $\tilde{\cal H}_{m}$ where the gauge group is confined 
down to $U(m)$. The upshot is a proposal for a new way of deconstructing 
Little String Theory as a large-$N$ limit of a confining gauge theory in four 
dimensions. The complete proposal is as follows;  
\paragraph{}
{\bf 1:} We consider the $U(N)$ 
$\beta$-deformed theory with renormalized gauge coupling 
$\tilde{g}_{R}^{2}$, 
deformation parameter $\tilde{\beta}=8\pi i/\tilde{g}_{R}^{2}n$ 
(with $n|N$) and zero vacuum angle. 
This theory has a Higgs branch where the $U(N)$ gauge group is confined 
down to a $U(m)$ subgroup. We consider the vacuum on this branch specified 
as in (\ref{chiralvevs1b},\ref{chiralvevs2b}) by complex numbers 
$\tilde{\alpha}_{1}$ and $\tilde{\alpha}_{2}$ with $\tilde{\alpha}_{3}=0$.   
\paragraph{}
{\bf 2:} We take the limit $N\rightarrow \infty$ with 
$m$ and $\tilde{g}^{2}_{R}$ held fixed, while scaling the VEVs as, 
\begin{eqnarray}
|\tilde{\alpha}_{1}| \sim \sqrt{n}\rightarrow \infty & \qquad \qquad & 
|\tilde{\alpha}_{2}| \sim \sqrt{n}\rightarrow \infty \nonumber \\
\label{scalings3q}
\end{eqnarray}  
These scalings leave the BPS string tension
$\tilde{T}=8\pi^{2}|\tilde{\alpha}_{1}||\tilde{\alpha}_{2}|/
\tilde{g}_{R}^{2}n$ fixed. 
\paragraph{}
{\bf 3:} Finally we take the zero coupling limit $\tilde{g}_{R}^{2}\simeq
\tilde{g}^{2}\rightarrow 0$ with the string tension $\tilde{T}$ held fixed. 
In this limit the world-volume non-commutativity disappears and the central 
$U(1)$ of the gauge group decouples. 
\paragraph{}
We propose that the resulting theory is the ${\cal N}=(1,1)$ 
Little String Theory with low-energy gauge group $SU(m)$. This theory has a 
single dimensionful parameter: the tension of the little string. This is 
identified with the BPS string tension $\tilde{T}$ of the 
four-dimensional gauge theory. 

\subsection{Another continuum limit}
\paragraph{}
We would also like to discuss a limit 
analogous to that considered in \cite{AHCK} which yields 
a commutative theory in six dimensions on a torus of fixed size. 
There is a unique way to take the lattice spacing to zero as 
$n\rightarrow \infty$ while holding $R_{1}$ and $R_{2}$ fixed,  
We must scale the VEVs as 
\begin{eqnarray}
|\alpha_{1}| \sim n\rightarrow \infty & \qquad \qquad & 
|\alpha_{2}| \sim n\rightarrow \infty \nonumber \\
\label{scalings3}
\end{eqnarray} 
This means that the non-commutativity parameter $\vartheta=
n/2\pi|\alpha_{1}||\alpha_{2}|\sim 1/n$ and thus goes to zero as required.   
If we also want to hold $G^{2}_{6}=
|\alpha_{1}||\alpha_{2}|/g^{2}n$ we are forced to take 
$g^{2}\rightarrow \infty$ holding $g^{2}/n$ fixed. 
\paragraph{}
Like the limit discussed in the previous section, this 
limit can be reinterpreted using the S-duality of the $\beta$-deformed 
theory. It is equivalent to 
a simultaneous $N\rightarrow \infty$, 
$\tilde{g}^{2}_{R}\rightarrow 0$ limit of a vacuum on the 
dual confining branch $\tilde{\cal H}_{m}$, taken holding the 
combination $\tilde{g}_{R}^{2}N$ and the vacuum moduli 
$\tilde{\alpha}_{1}$, $\tilde{\alpha}_{2}$ fixed.  Hence the BPS string 
tension 
$\tilde{T}$ given in (\ref{t2}) also remains constant 
in the limit. In fact this is just a standard 't Hooft limit of the confining 
gauge theory in a fixed vacuum state on the branch $\tilde{\cal H}_{m}$.      
\paragraph{}
In \cite{nd2}, we present evidence that the resulting continuum theory 
is again commutative ${\cal N}=(1,1)$ Little String Theory with 
string tension identified 
with the BPS string tension $\tilde{T}$. The new feature 
is that the Little String Theory is compactified to 
four dimensions on a torus of fixed radii $R_{1}$ and $R_{2}$.

\subsection{Deconstructing the Coulomb branch}
\paragraph{}
In the preceeding two subsections we have proposed limits 
where the dynamics of the $\beta$-deformed theory is described by 
Little String Theory. In particular, we focused on the large-$N$ limit of 
the theory on confining branches where the unconfined subgroup of the gauge 
group is $U(m)$. In these limits the central $U(1)$ subgroup decouples 
and we obtain the Little String Theory on $m$ coincident NS5 branes 
whose low-energy gauge group is $SU(m)$. However, Little String Theory also 
has a Coulomb branch where the NS5-branes are separated breaking the 
low-energy gauge group to $U(1)^{m-1}$. In this phase, 
the spectrum also contains `W-bosons' corresponding to D-strings stretched 
between the fivebranes. In particular one can define a double-scaling 
limit \cite{GK}, where the W-boson masses are kept fixed in the NS fivebrane 
decoupling limit $\tilde{g}_{s}\rightarrow 0$, with $\tilde{\alpha}'$ 
kept fixed as before. The resulting theory has a holographic dual which 
includes six flat directions and an internal space which is 
(roughly speaking) the product of a 
compact coset and the two-dimensional Euclidean black-hole \cite{sf}. 
Importantly, as the 
starting point involved only Neveu-Schwarz branes, there are no 
background Ramond-Ramond fields. When the W-boson mass is much larger 
than the scale set by the string tension, the dilaton is everywhere 
small and it suffices to consider tree-level string theory in this 
background. The corresponding worldsheet theory is solvable and and 
the correlation functions can be calculated explicitly 
\cite{GK}. 
\paragraph{}
The two proposed continuum limits described above can easily be modified 
to give Little String Theory in its Coulomb phase. To accomplish this 
we simply need to move onto the Coulomb branch of the low-energy 
$U(m)$ theory.  This corresponds to a phase discussed above where the 
$U(N)$ gauge group of the $\beta$-deformed theory 
is confined down to $U(1)^{m}$. In this case the BPS spectrum of the 
low-energy theory will also include W-bosons. 
We can now take the same limits described in Sections 9.1 and 9.2 with 
the ratio $x=M_{W}/\sqrt{\tilde{T}}>>1$ held fixed where 
$M_{W}$ is the mass of the lightest magnetic W-boson and, as above, 
$\tilde{T}$ is the string tension. This is particularly 
clear in the string theory construction of Section 8, where the 
corresponding configuration involves $m$ NS5 branes wrapped on concentric 
torii of different radii. As in the set up of \cite{GK} the W-bosons 
correspond to D-strings stretched between NS5 branes. In the present case, 
their masses 
are controlled by the differences in radii of the toroidally-wrapped 
fivebranes. Each radius 
corresponds to a modulus of the configuration which can be 
varied independently\footnote{For example, one can adjust the 
VEVs so that the masses of W-bosons match those found in the vacuum  
configuration described in Eq (1.10) of the second reference in \cite{GK}.}.   \paragraph{}
By taking the limit discussed in Section 9.2 while keeping the W-boson 
mass fixed we arrive at a proposal for deconstructing double-scaled Little 
String Theory compactified on a torus of fixed size. 
The starting point is a 't Hooft large-$N$ limit of a four-dimensional 
gauge theory in a phase where the 
$U(N)$ gauge group is confined down to 
a product of $U(1)$'s. As discussed in Section 7, the theory on this branch 
has a semiclassical regime where we can 
explicitly exhibit confinement of electric charges due to the formation of 
BPS flux tubes of fixed core size. In this regime the scale set by the 
string tension is much larger than the W-boson mass.  
In the previous paragraph we have described another 
regime of the large-$N$ theory on this branch where the strings become 
light compared to the W-bosons. The effective description of the new regime is 
given by double-scaled Little String Theory compactified on a torus. 
The correlation functions of double-scaled Little String Theory 
calculated in \cite{GK} are consistent with the 
existence of an infinite tower of stable, 
higher-spin states which behave like free particles when $x>>1$. 
This matches well with our expectations for the large-$N$ 
glueball spectrum of a confining gauge theory containing 
only adjoint fields\footnote{In fact, the similarity between the spectrum of 
double-scaled Little String Theory and that of a large-$N$ gauge theory was 
noted in \cite{GK}}. The $\beta$-deformed theory therefore seems to 
provide a new and surprisingly tractable version of the duality 
between large-$N$ confining gauge fields and weakly 
interacting strings \cite{TH1}. This is discussed further in \cite{nd2}. 
\paragraph{}
The author would like to thank Ofer Aharony, Prem Kumar and 
Richard Szabo for their useful 
comments on a preliminary version of this paper. 
I would also like to thank Tim Hollowood for 
discussions. 

\section*{Appendix A: S-duality properties of the 
$SU(N)$ and $U(N)$ Theories}
\paragraph{}
In the $\beta=0$ case the 
difference between the theories with gauge groups $SU(N)$ and 
$U(N)$ is trivial. When $\beta\neq 0$, the situation is much more 
complecated because the additional trace fields of the $U(N)$ theory,  
denoted $a_{i}={\rm Tr}_{N} \Phi_{i}/N$ for $i=1,2,3$, are now coupled to the 
traceless $SU(N)$ fields. The results of \cite{DHK} take the simplest form 
for the $SU(N)$ theory: the exact superpotential in the massive 
vacua was given in Eq (\ref{resultdhk}) above. This expression 
(and the resulting condensates) transforms 
with a fixed modular weight under the modified $SL(2,{\bf Z})$ transformations 
(\ref{sl2z}). Thus we 
have a direct generalization of the $SL(2,{\bf Z})$ duality of the 
${\cal N}=4$ theory. 
\paragraph{}
The corresponding situation in the $U(N)$ theory is less 
straightforward. The $U(N)$ effective superpotential itself does 
not have simple transformation properties but instead the combination, 
\begin{equation}
\hat{\cal W}^{U(N)}_{\rm eff}= \frac{{\cal W}_{\rm eff}^{U(N)}}{1-\frac{4\kappa^{2}
\sin^{2}\left(\frac{\beta}{2}\right)}{N\mu M^{2}}{\cal W}_{\rm eff}^{U(N)}}
\label{what}
\end{equation}   
transforms in the same way as the $SU(N)$ superpotential\footnote{In fact 
they are equal: $\hat{\cal W}^{U(N)}_{eff}={\cal W}^{SU(N)}_{eff}$}. 
Operators with good modular properties can be defined by differentiating
$\hat{\cal W}_{\rm eff}$ with respect to parameters. For example 
we can define an operator $\hat{O}$ with     
\begin{equation}
\langle \hat{O} \rangle= 
\frac{\partial}{\partial \mu} \, \hat{\cal W}^{U(N)}_{\rm eff} 
\label{o2}
\end{equation}
which transforms with modular weight $(1,1)$, just as 
${\rm Tr}_{N} \Phi^{2}_{3}$ does in the $SU(N)$ theory.
The difference between the two expressions is subleading in either the 
large-$N$ or small-$|\beta|$ regimes which are the main 
concern of this paper. The modular behaviour of the two theories is the 
same in these regimes. This matches the results of the 
linearized analysis given in Section 2 which applies equally to 
gauge groups $U(N)$ and $SU(N)$.  
\paragraph{}
Away from the regime of large-$N$ or small $|\beta|$, it seems that the 
operators transforming with good modular weights are different for the 
$U(N)$ and $SU(N)$ theories. This complication arises because 
the trace field $a_{3}$ of the massive $U(N)$ theory, gets an 
expectation value in each vacuum \cite{DHK}. To be precise we have, 
\begin{equation}
\langle a_{3} \rangle= \frac{2\kappa\sin\left(\frac{\beta}{2}\right)}
{NM\mu}\, {\cal W}^{U(N)}_{\rm eff} 
\label{tracevev}
\end{equation}                 
\paragraph{}
Taking (\ref{what},\ref{o2},\ref{tracevev}) together, we find, 
\begin{equation}
\langle \hat{O} \rangle = 
\frac{\langle {\rm Tr}_{N}\Phi_{3}^{2} \rangle^{2}}{\left(\langle 
{\rm Tr}_{N} \Phi_{3}^{2}
 \rangle^{2}-\frac{1}{N^{2}}\langle {\rm Tr}_{N} \Phi_{3} \rangle^{2}\right) } 
\label{reparam}
\end{equation}
Similar relations 
hold for other condensates. Thus, we see that the operators $\hat{O}$ 
and ${\rm Tr}_{N} \Phi^{2}_{3}$ will have very different transformation 
properties whenever $a_{3}\neq 0$. On the other hand, if we can 
consistently set $a_{3}=0$ then the modular transformation properties of the 
$U(N)$ and $SU(N)$ theories will be the same. In the massless 
$\beta$-deformed theory considered in this paper $\langle a_{3} \rangle$ 
vanishes in all supersymmetric vacuum states. For this reason we believe 
that the modular properties of the {\em massless} $\beta$-deformed theories 
with gauge groups $U(N)$ and $SU(N)$ are actually the same.        

\section*{Appendix B: Classical branches of the massive theory}
\paragraph{}
In this Appendix we consider massive version of the $\beta$-deformed theory. 
We first consider introducing a non-zero mass $\mu$ for a single chiral 
multiplet. This corresponds to seting $M=0$ in (\ref{rel}) which gives  
$\Delta{\cal W}=\mu{\rm Tr}_{N}\Phi^{2}_{3}$. Perhaps 
surprisingly, the classical Higgs branch ${\cal H}_{1}$ which occurs 
for $\beta=2\pi /N$ is not lifted by this 
perturbation\cite{BJL1}. Setting,  
\begin{equation}
\begin{array}{ccc} \langle \Phi_{1}\rangle = \alpha_{1}U_{(N)} & \qquad{} 
\qquad{} 
\langle  \Phi_{2}\rangle = \alpha_{2}V_{(N)}+
\gamma V_{(N)}^{\dagger}U_{(N)}^{\dagger\, 2} & \qquad{} \qquad{}  
\langle \Phi_{3}\rangle = \alpha_{3}W_{(N)} \end{array}
\label{ansatz}
\end{equation} 
we find that the modified F-term vacuum equations are satisfied provided 
$\alpha_{1}\gamma \kappa\sin(\pi/N)=\mu\alpha_{3}$. This configuration is 
not D-flat, but can be made so by a complex gauge transformation. 
Thus we see that a 
three-parameter family of vacua survives for $\mu\neq 0$. As the matrices 
appearing in (\ref{ansatz}) are traceless this statement is equally true 
for gauge groups $U(N)$ and $SU(N)$. This construction generalises easily to 
the other Higgs branches ${\cal H}_{m}$ for $m>1$.  Generic  
vacua have non-zero VEVs for the chiral operators $x$, $y$, $z$ and $w$, 
but the equation (\ref{orb1}) is modified. This is consistent because 
introducng $\mu$ explicitly breaks one of the $U(1)$ R-symmetries of the 
$\beta$-deformed theory. In fact the modified equation describes a resolution 
of the orbifold, where one of the three fixed lines described above 
has been eliminated.     
\paragraph{}    
As in the massless theory, acting with the S-duality transformation 
(\ref{sl2z}) produces an $SL(2,Z)$ multiplet of new branches. As discussed 
in Section 7, the fact that 
these vacua are still present in the theory with $\mu\neq 0$ is a non-trivial 
check on our discussion of the massless limit.
\paragraph{}
We now turn on both mass parameters $M$ and $\mu$. In this case the 
behaviour of the $U(N)$ and $SU(N)$ theories is different. The classical 
vacuum structure of both theories can be deduced from the results of 
\cite{BJL1} and \cite{DHK}. In the 
$SU(N)$ theory, the branch (\ref{ansatz}) is lifted while a continuous 
degeneracy of vacua survives in the $U(N)$ theory. Moving away from 
$\beta=2\pi/N$ both theories only have isolated classical 
vacua corresponding to deformed $SU(2)$ representations.  One can 
explicitly check that, in the $SU(N)$ theory, one of these isolated 
vacua approaches a point on the Higgs branch (\ref{ansatz}) as we take 
the $M\rightarrow 0$, $\beta\rightarrow 2\pi/N$ limit discussed in 
the text. In contrast, the corresponding vacuum of the $U(N)$ theory 
instead approaches the root of this branch, a point with 
$\langle w \rangle=0$, in this limit.

\end{document}